\documentclass[10pt,journal]{IEEEtran}
\usepackage{amsfonts}
\usepackage{amssymb,epic,eepic}
\usepackage{amsmath}

\newcommand{\hr}{{\cal H}}

\newcommand{\cc}{{\mathbb C}}

\newcommand{\nn}{{\mathbb N}}
\newcommand{\idn}{\mathbf{1}}
\newcommand{\zz}{{\mathbb Z}}

\newcommand{\eps}{{\varepsilon}}        

\newtheorem{theorem}{Theorem}[section]         
\newtheorem{lemma}[theorem]{Lemma}             
\newtheorem{corollary}[theorem]{Corollary}     
\newtheorem{remark}[theorem]{Remark}           
\newtheorem{proposition}[theorem]{Proposition} 

\begin{document}

\title{Ergodic Classical-Quantum Channels: Structure and Coding Theorems}

\author{Igor Bjelakovi\'c and Holger Boche \IEEEmembership{Member, IEEE}%
\thanks{This work is supported by the Deutsche Forschungsgemeinschaft DFG via project Bj 57/1-1 ''Entropie und Kodierung gro\ss er Qunten-Informationssysteme''.}
\thanks{The authors are with the Heinrich-Hertz-Chair for Mobile Communications, Technische Universit\"at Berlin, Werner-von-Siemens-Bau (HFT 6), Einsteinufer 25, 10587 Berlin, Germany and the Institut f\"ur Mathematik, Technische Universit\"at Berlin, Stra\ss e des 17. Juni 136, 10623 Berlin, Germany}
}

\maketitle
\begin{abstract}
We consider ergodic causal classical-quantum channels (cq-channels) which additionally have a decaying input memory. In the first part we develop some structural properties of ergodic cq-channels and provide equivalent conditions for ergodicity. In the second part we prove the coding theorem with weak converse for causal ergodic cq-channels with decaying input memory. Our proof is based on the possibility to introduce a joint input-output state for the cq-channels and an application of the Shannon-McMillan theorem for ergodic quantum states. In the last part of the paper it is shown how this result implies a coding theorem for the classical capacity of a class of causal ergodic quantum channels. 
\end{abstract}

\section{Introduction}
One of the main achievements in quantum information theory is the determination of the capacity of memoryless quantum channels by Holevo \cite{holevo-1, holevo-2}, and independently by Schumacher and Westmoreland \cite{schumacher}, for transmission of classical information. These results have been considerably sharpened by Winter \cite{winter}, who extended Wolfowitz's \cite{wolfowitz} approach via frequency typical sequences to the quantum setting and obtained coding theorem with strong converse for transmission of classical information over memoryless quantum channel. At the same time Ogawa and Nagaoka \cite{ogawa-nagaoka} proved the strong converse in the memoryless situation by a different proof that follows the classic Arimoto's \cite{arimoto} approach. 
Subsequently Shor \cite{shor} and Devetak \cite{devetak} have shown by independent proofs that  the capacity of the memoryless quantum channel is given by the coherent information. The weak converse to this coding theorem was already established by Barnum, Nielsen and Schumacher in \cite{barnum-nielsen-schumacher}.
 Shor uses the method of random selection of subspaces while Devetak chooses an approach via private classical capacity and a transformation of private classical codes into quantum codes. An interesting point in Devetak's approach is that the classical capacity results for quantum channels \cite{holevo-2, schumacher, winter} are one of the crucial building blocks for the direct part of coding theorem for quantum channels. In spite of this progress one of the main open questions concerning classical capacity of memoryless quantum channels, namely the additivity problem \cite{shor-2}, is still unresolved.\\
Although memoryless quantum channels play a prominent role in the development of the foundations of quantum information theory, real-world channels are rarely memoryless. Thus it is desirable to have a broad and efficient theory for quantum channels with memory.  
In this paper we will consider causal ergodic classical-quantum channels satisfying an additional continuity condition, which basically means that the effect of the inputs far in the past decays with respect to the variational distance. The causality means in this context that the outputs up to time $t$ depend only on inputs up to time $t$. Obtained coding results are then applied to a class of causal quantum channels in order to obtain their classical capacity. The extension to causal ergodic quantum channels with decaying input memory is not obvious at all, since several ergodicity problems arise. The discussion of these problems is postponed to the future work.  
\subsection{Overview and Outline}
In section \ref{cq-channels} we introduce general classical-quantum channels (cq-channels for short) in analogous fashion to the classical setting as a family of conditional states and show how this point of view leads to the usual definition as a completely positive map between the output-algebra and input-algebra.
Concepts of stationarity and ergodicity are also introduced and an equivalent condition for ergodicity of a stationary cq-channel is derived. It states that a stationary cq-channel is ergodic iff it is extreme point in the set of all stationary cq-channels. Several equivalent formulations of this statement are given and it is shown that not all of them can be extended to stationary quantum channels simultaneously.\\
The second part of section \ref{cq-channels} is devoted to the notion of \emph{continuity of cq-channels}. As it was pointed out by McMillan in his classic paper \cite{mcmillan} the continuity properties of channels are crucial for the validity of coding theorems. We formulate these continuity notions with respect to the \emph{variational distance} on quantum states. A more natural notion of distance would be a metric on quantum states which shares as many properties as possible with the classical $\bar{d}-$distance, which is extremely sensitive to the ergodic and mixing properties of channels and probability measures. See e.g. Gray and Ornstein \cite{gray-ornstein} for application to classical channels, and books \cite{shields, gray} by Shields resp. Gray for introduction and applications to the ergodic and information theory. To our knowledge, there is still no metric on the set of quantum states which could play a similar role as the $\bar{d}-$distance.\\
Section \ref{cq-channels} is closed by some examples.\\
After some preliminary results in section \ref{preliminary} we prove the direct coding theorem and weak converse to it in section \ref{coding-imc}. Our approach combines the maximal code construction of Wolfowitz \cite{wolfowitz}, which was already applied by Winter \cite{winter} for memoryless cq-channels, with the quantum Shannon-McMillan theorem from \cite{bksss}, in order to avoid usage of frequency typical and conditionally frequency typical projections, which are not an adequate tool for correlated quantum states. In a sense, this approach is a mixture of Wolfowitz's code construction and the version of Feinstein's lemma from Blackwell/Breiman/Thomasian in \cite{blackwell} which is based on the notion of the joint input-output probability distribution.\\
These results are extended to causal cq-channels with decaying input memory in section \ref{coding-dima}. Since our notion of continuity is symmetric with respect to past and future it is even possible to prove coding theorem for cq-channels with decaying input memory \emph{and} anticipation without any additional complications, although we are only interested in causal situations.\\
Finally, the results are applied to obtain coding results for transmission of classical information via weakly output mixing quantum channels in section \ref{coding-class-quant}. In order to keep the  extent of this paper reasonable, the extension to ergodic quantum channels is postponed to a forthcoming paper.\\
For a reader without experience with $C^{\ast}-$algebras, states and quasi-local algebras we provide an appendix that contains a short description of these concepts complemented by some standard references and some examples.
\subsection{Related Work}
In \cite{kretschmann-werner} Kretschmann and Werner considered stationary causal quantum channels, and have shown that each channel of this type can be seen as a concatenation of channels acting on the single-site input algebra and memory algebra with the range in the joint system of the single-site output algebra and memory algebra. In the classical setting this approach corresponds to the point of view, that the past inputs and outputs can be seen as the states of the channel for the transmission of the actual input letter and the resulting output symbol. If the duration of the memory of the input and output is finite this class of channels is called \emph{finite-state channels} and coding theorems for them were established by Blackwell, Breiman and Thomasian in \cite{blackwell-1} under the assumption that the channel is additionally \emph{indecomposable}. A different account to finite-state indecomposable channels is given in the monograph \cite{gallager} by Gallager.\\
Kretschmann and Werner have proved in \cite{kretschmann-werner} a general weak converse to the coding theorem, and for channels with \emph{finite memory} they derived the direct part of the coding theorem under the assumption that the channel is \emph{forgetful}, which is quantum analogon of the notion of the indecomposability. Under these circumstances the channels under consideration are well approximable by memoryless quantum channels.\\
Recently, Datta and Dorlas \cite{datta-dorlas} used an approach via a quantum version of Feinstein's Lemma (cf. \cite{feinstein}) to give another proof of the direct part of the coding theorem for transmission of classical information over a quantum memoryless channel and remarked that this approach can be extended to prove direct coding theorem for classical information transmission via quantum channels with Markovian correlated noise (see Section \ref{examples} for the definition of this class of channels). This class of channels was introduced by Macchiavello and Palma in \cite{macchiavello} and bounds for capacity were already obtained by Bowen and Mancini in \cite{bowen-mancini}.

\subsection{Notation}
In this paper we will write $\mathfrak{F}(Y)$ for the set of $\cc -$valued functions defined on a finite set $Y$. The set of linear operators (linear maps) acting on a finite-dimensional Hilbert space $\hr$ will be denoted by $\mathcal{L}(\hr)$ and this set will be often abbreviated by $\mathcal{B}$. $[n,k]$ for integers $n,k$ with $n\le k$ stands for integer intervals, i.e. the set of all $z\in\zz$ that satisfy $n\le z\le k$.\\
For a given probability measure $p$ on a measurable space $(\Omega,\Sigma)$ we will not distinguish between the the measure and the expectation functional generated by it, i.e. for an integrable $\cc -$valued function $f$ on $\Omega$
we set
\[p(f):=\int f(\omega)p(d\omega). \] 
Notation concerning $C^{\ast}-$algebras, quasi-local algebras $\mathcal{B}^{\zz}$ built up from $\mathcal{B}$, and states is introduced in the appendix. Moreover, the appendix contains the definitions of stationary and ergodic states on quasi-local algebras and von Neumann entropy rate of stationary states.\\
For a finite set $A$ or a finite dimensional $C^{\ast}-$algebra $\mathcal{B}$, $A^n$ and $\mathcal{B}^{n}$,$ n\in\nn$, stand for $A^{[1,n]}$ resp. $\mathcal{B}^{[1,n]}$. Restriction of a state $\psi$ on $\mathcal{B}^{\zz}$ to $\mathcal{B}^n$ is denoted by $\psi^n$; $\textrm{tr}$ denotes the trace of operators. 
\section{Classical-Quantum Channels}\label{cq-channels}
Let $A$ be a finite set and let $\hr$ be a $d$-dimensional Hilbert space. By $A^{\zz}$ we denote the set of doubly infinite sequences with components from $A$ and $\mathcal{B}^{\zz}$ is the quasi-local $C^{*}$-algebra with $\mathcal{B}:=\mathcal{L}(\hr)$ (cf. appendix \ref{quasi-local-app}). Moreover, let $\cal{S}(\mathcal{B}^{\zz})$ denote the set of \emph{states} on $\mathcal{B}^{\zz}$, i.e. the set of positive linear normalized functionals on $\mathcal{B}^{\zz}$ with values in $\cc$. Since $A^{\zz}$ can be easily equipped with a metric making it a compact space, we have a natural notion of Borel $\sigma$-field which coincides with the $\sigma$-field generated by the cylinder sets denoted by $\Sigma_{c}$. Moreover all cylinder sets are open as well as closed with respect to this topology. One possibility to introduce such a metric on $A^{\zz}$ is
\[d(x,y):=\sum_{i\in\zz}2^{-|i|}d_{H}(x_i,y_i)\qquad (x,y\in A^{\zz}),\]
where $d_{H}$ denotes familiar Hamming distance. The properties of the resulting topology mentioned above are then standard. The set of bounded Borel measurable complex-valued functions will be denoted by $B(A^{\zz},\Sigma_{c})$ and it is always assumed that this set is endowed with the $||\cdot ||_{\infty}$-norm. Note that $B(A^{\zz},\Sigma_c)$ is a commutative $C^{\ast}$-algebra when equiped with $||\cdot||_{\infty}$ and complex-conjugation as adjoint operation (cf. appendix \ref{c-ast-app} for definition).\\
We consider a \emph{classical-quantum channel} (cq-channel) $W$ with the input $A^{\zz}$ and output $\mathcal{B}^{\zz}$, i.e. a map $W:A^{\zz}\times \mathcal{B}^{\zz}\to \cc$ with following properties
\begin{enumerate}
\item For each $b\in \mathcal{B}^{\zz}$ the map $x\mapsto W(x,b)$ is Borel measurable.
\item For each $x\in A^{\zz}$ the map $b\mapsto W(x,b)$ is a state.
\end{enumerate}
\begin{remark}\label{rem-bounded-borel}
 The fact that the map $b\mapsto W(x,b)$ is a state for all $x\in A^{\zz}$ implies that $|W(x,b)|\le ||b||$ holds and hence that the first item above can be sharpened to the statement that $x\mapsto W(x,b)$ is \emph{bounded} Borel function.
\end{remark}
Note that each cq-channel $W:A^{\zz}\times \mathcal{B}^{\zz}\to \cc $ defines a linear (completely) positive unital map $K:\mathcal{B}^{\zz}\to B(A^{\zz},\Sigma_{c}) $ given by $(K(b))(x):=W(x,b)$. Conversely, each linear map $K:\mathcal{B}^{\zz}\to B(A^{\zz},\Sigma_{c}) $ with properties mentioned above gives rise to a cq-channel via $(K(b))(x)=:W(x,b)$. Consequently we have a one-to-one affine correspondence between linear positive unital maps and cq-channels. Hence the definition of the channel as a completely positive unital map between $C^{*}$-algebras is recovered in this situation. Moreover, if the input or output algebra is abelian the mere positivity is sufficient, since in this situation each positive linear map is automatically completely positive (cf. \cite{stinespring, paulsen}). We summarize this trivial observation to ease later referencing in
\begin{lemma}\label{trivial-1}
There is a one-to-one affine correspondence between cq-channels $W:A^{\zz}\times \mathcal{B}^{\zz}\to \cc $ and linear positive unital maps $K:\mathcal{B}^{\zz}\to B(A^{\zz},\Sigma_{c}) $ which is given by
\begin{equation}\label{k-vs-w} 
(K(b))(x)=W(x,b).
\end{equation}
\end{lemma} 
The joint \emph{input-output state} on $B(A^{\zz},\Sigma_{c})\otimes \mathcal{B}^{\zz} $ is defined by linear extension of
\begin{equation}\label{joint-state}
\psi_{p,W}(1_{C}\otimes b):=\int_{C}W(x,b)p(dx)\quad (C\in \Sigma_{c}, b\in \mathcal{B}^{\zz}),
\end{equation}
for a given input probability measure $p$ on $(A^{\zz},\Sigma_{c})$. Note that the integral above is well defined due to assumed measurability of the map $A^{\zz}\ni x\mapsto W(x,b) $ for each $b\in \mathcal{B}^{\zz}$. It is clear that $\psi_{p,W}(1_{C}\otimes \idn)=p(C)$, and now we compute 
\begin{eqnarray*}
\psi_{p,W}(\idn\otimes b)&=&\int_{A^{\zz}}W(x,b)p(dx)\\ 
&=&\int_{A^{\zz}} (K(b))(x)p(dx)\quad (\textrm{by Lemma \ref{trivial-1}})\\
&=& p(K(b)),
\end{eqnarray*}
which is the correct formula for the output state.\\
The formula in eq. (\ref{joint-state}) can be naturally extended to bounded Borel functions instead of indicator functions $1_{C}$ by setting
\begin{equation}\label{joint-state-2}
 \psi_{p,W}(f\otimes b)=\int f(x)W(x,b)p(dx)
\end{equation} 
for $f\in B(A^{\zz},\Sigma_c), b\in \mathcal{B}^{\zz}$. We establish that (\ref{joint-state}) defines a state after the following
\begin{remark}\label{rem-tensor-product}
 Note that in general for two $C^{*}$-algebras $\mathcal{A}$, $\mathcal{B}$ there are several norms on $\mathcal{A}\otimes \mathcal{B}$ making it $C^{*}$-algebra. However, if one of the factors is nuclear (as are abelian, finite-dimensional and quasi-local $C^{*}$-algebras) there is a unique $C^{*}$-norm on $\mathcal{A}\otimes \mathcal{B}$ (see \cite{kadison-ringrose, paulsen} for details).
\end{remark}
\begin{lemma}\label{joint-state-is-state}
The functional defined by linear extension of $\psi_{p,W}$ in eq. (\ref{joint-state-2}) is a state on $B(A^{\zz},\Sigma_c)\otimes \mathcal{B}^{\zz}$.
\end{lemma}
\begin{proof} For a given cq-channel $W:A^{\zz}\times \mathcal{B}^{\zz}\to \cc$ let us consider the corresponding positive, unital linear map $K:\mathcal{B}^{\zz}\to B(A^{\zz},\Sigma_c)$ from Lemma \ref{trivial-1}. Let us define the copier $Copy: A^{\zz}\to A^{\zz}\times A^{\zz}$, $Copy(x):=(x,x)$, and the induced map $Copy: B(A^{\zz},\Sigma_c)\otimes B(A^{\zz},\Sigma_c)\to B(A^{\zz},\Sigma_c)$ which is given by $Copy(f):=f\circ Copy$ (Here we identify $B(A^{\zz},\Sigma_c)\otimes B(A^{\zz},\Sigma_c) $ with $B(A^{\zz}\times A^{\zz},\Sigma_c\times \Sigma_c )$). Note that this last map is linear, positive and unital, and thus completely positive and unital (see discussion preceding Lemma \ref{trivial-1}). Moreover the map $\textrm{id}_{B(A^{\zz},\Sigma_c)}\otimes K:B(A^{\zz}, \Sigma_c)\otimes \mathcal{B}^{\zz}\to B(A^{\zz}, \Sigma_c)\otimes B(A^{\zz}, \Sigma_c)  $ is completely positive and unital (see \cite{paulsen} chap. 12). Let us define $E:B(A^{\zz}, \Sigma_c)\otimes \mathcal{B}^{\zz}\to B(A^{\zz}, \Sigma_c) $ by
\begin{equation}\label{cp-map-joint-channel}
 E:=Copy\circ (\textrm{id}_{B(A^{\zz},\Sigma_c)}\otimes K ).
\end{equation}
As composition of completely positive unital maps E is itself completely positive and unital. Thus 
\[(p\circ E)(a)=\int (E(a))(x)p(dx) \quad (a\in B(A^{\zz},\Sigma_c)\otimes \mathcal{B}^{\zz} )\]
defines a state on $B(A^{\zz},\Sigma_c)\otimes \mathcal{B}^{\zz}$, and it is apparent that this state coincides with $\psi_{p,W}$ on elementary tensors $f\otimes b$ and (finite) linear combinations thereof. Consequently they must coincide at all.\end{proof}
Note that eq. (\ref{cp-map-joint-channel}) sets up a one-to-one affine correspondence between linear, positive unital maps $K:\mathcal{B}^{\zz}\to B(A^{\zz},\Sigma_c) $ and linear positive unital maps $E:B(A^{\zz}, \Sigma_c)\otimes \mathcal{B}^{\zz}\to B(A^{\zz}, \Sigma_c) $ with
\begin{equation}\label{factorization}
 E(f\otimes b)=fE(\idn_{B(A^{\zz},\Sigma_c)}\otimes b)=fK(b),
\end{equation} 
as is easily verified. Thus combining this simple observation and Lemma \ref{trivial-1} we obtain
\begin{theorem}\label{affine-corresp-of-all-things}
There is one-to-one
 affine correspondence between
 \begin{enumerate}
 \item cq-channels $W:A^{\zz}\times \mathcal{B}^{\zz}\to \cc$,
\item linear, positive unital maps $K:\mathcal{B}^{\zz}\to B(A^{\zz},\Sigma_c) $ and
\item linear, positive unital maps $E:B(A^{\zz}, \Sigma_c)\otimes \mathcal{B}^{\zz}\to B(A^{\zz}, \Sigma_c) $ with
 \[ E(f\otimes b)=fE(\idn\otimes b).\]
\end{enumerate}
The correspondences are given by eq. (\ref{k-vs-w}) and eq. (\ref{cp-map-joint-channel}).
\end{theorem}
We emphasize at this point that these results heavily depend on the fact, that the input algebra $B(A^{\zz}, \Sigma_c)$ is abelian (commutative). This can be seen in eq. (\ref{cp-map-joint-channel}) where we used the universal copier $Copy$, a device which is provably impossible to introduce in a truly quantum mechanical setting. (see \cite{zurek, maassen, werner} for various versions of this fact known as No-Cloning theorem). That the affine correspondence in Theorem \ref{affine-corresp-of-all-things} cannot be valid for general channels (i.e. completely positive unital maps $K:\mathcal{B}\to \mathcal{A}$ between general $C^{\ast}-$algebras) can be seen from following theorem called Heisenberg's principle in \cite{maassen} , which states that for each completely positive unital map which would give rise to a joint input-output state the corresponding channel ($K(b)=E(\idn\otimes b)$ in our case) has necessarily commutative range:
\begin{theorem}[Heisenberg's Principle \cite{maassen}]\label{heisenberg}
Let $\mathcal{A},\mathcal{B}$ be arbitrary unital $C^{\ast}-$algebras and let $E:\mathcal{A}\otimes \mathcal{B}\to \mathcal{A}$ be a completely positive unital map with
\[E(a\otimes\idn_{\mathcal{B}} )=a \quad \forall a\in\mathcal{A}. \]
Then
\[ E(\idn_{\mathcal{A}}\otimes b)\in \mathcal{Z}(\mathcal{A})\quad \forall b\in \mathcal{B}, \]
where $\mathcal{Z}(\mathcal{A}) $ denotes the center of $\mathcal{A}$ given by
\[\mathcal{Z}(\mathcal{A}):=\{ a'\in\mathcal{A}: a'a=aa'\quad \forall a\in\mathcal{A}\},  \]
which is an abelian $^{\ast}-$subalgebra of $\mathcal{A}$.
\end{theorem}
The proof of this theorem in \cite{maassen} is given for finite-dimensional algebras, but it can be immediately extended to arbitrary $C^{\ast}-$algebras, since all necessary ingredients are valid generally (cf. \cite{paulsen}). Thus, in general situation we have merely a completely positive unital map $K:\mathcal{B}\to\mathcal{A}$ as the description of the quantum channel at our disposal.\\
A cq-channel $W$ is called \emph{stationary} if $W(T_{\textrm{in}}x,b)=W(x,T_{\textrm{out}}b)$ holds for all $x\in A^{\zz}$ and all $b\in \mathcal{B}^{\zz}$, where $T_{\textrm{in}}$ resp. $T_{\textrm{out}}$ denotes the shift on the input alphabet resp. output algebra. By virtue of Lemma \ref{trivial-1} this is equivalent to $K\circ T_{\textrm{out}}=T_{\textrm{in}}\circ K$.\\
A cq-channel $W$ is called \emph{ergodic}, if it is stationary and if the joint input-output state $\psi_{p,W}$ is ergodic for every stationary ergodic probability measure $p$ on $(A^{\zz},\Sigma_{c})$. \\
\subsection{Structural Properties of Ergodic CQ-Channels}
At this point we pause with further definitions and give an alternative characterization of ergodicity of cq-channels which parallels the characterization of ergodic states and probability measures. To this end we need the notion of equality of two cq-channels $W_1$ and $W_2$: Stationary cq-channels $W_1,W_2:A^{\zz}\times \mathcal{B}^{\zz}\to\cc $ are equal if for all $f\in B(A^{\zz},\Sigma_c), b\in \mathcal{B}^{\zz}$ and all stationary $p\in \mathcal{P}(A^{\zz},\Sigma_c )$
\begin{equation}\label{channel-equality-1}
 \psi_{p,W_1}(f\otimes b)=\psi_{p,W_2}(f\otimes b)
\end{equation}
holds, i.e. if $W_1$ and $W_2$ generate the same stationary joint input-output states. Note that this is equivalent to the assertion that for all $b\in\mathcal{B}^{\zz}$ and all stationary probability measures $p$ we have $W_1(x,b)=W_2(x,b)$ almost surely with respect to $p$. In view of Theorem \ref{affine-corresp-of-all-things} this can be rephrased by
\begin{equation}\label{channel-equality-2}
 p(fK_{1}(b))=p(fK_{2}(b)),
\end{equation}
or
\begin{equation}\label{channel-equality-3}
 p(E_1(f\otimes b))=p(E_2(f\otimes b))
\end{equation}
for all $f\in B(A^{\zz},\Sigma_c), b\in \mathcal{B}^{\zz} $ and all stationary $p\in \mathcal{P}(A^{\zz},\Sigma_c ) $.\\

\begin{theorem}\label{erg-vs-extremal}
Let $W: A^{\zz}\times \mathcal{B}^{\zz}\to\cc$ a stationary cq-channel. Then following assertions are equivalent:
\begin{enumerate}
\item[1)] $W$ is extremal in the convex set of stationary cq-channels
\item[2)] $W$ is ergodic
\end{enumerate}
\end{theorem}
\begin{proof} 2) $\Rightarrow$ 1). Suppose that $W$ is not extremal. Then there are stationary cq-channels $W_1 , W_2 : A^{\zz}\times\mathcal{B}^{\zz}\to\cc $, $W_1\neq W_2$, and $a\in (0,1)$ with
\begin{equation}\label{conv-decomp}
W=a W_1 + (1-a)W_2.
\end{equation}
Since $W_1\neq W_2$ there is a stationary $p\in \mathcal{P}(A^{\zz},\Sigma_c ) $ and $f\in B(A^{\zz},\Sigma_c), b\in \mathcal{B}^{\zz} $ with
\[ \psi_{p,W_1}(f\otimes b)\neq\psi_{p,W_2}(f\otimes b).\]
Note that applying the ergodic decomposition of $p$ (cf. \cite{gray-davisson} or \cite{shields} Sec I.4) we may assume that $p$ is already stationary ergodic. Then using eq. (\ref{conv-decomp}) we obtain a convex decomposition of the joint input-output state for $W$:
\[\psi_{p,W}=a \psi_{p,W_1}+ (1-a)\psi_{p,W_2}, \]
with $\psi_{p,W_1}(f\otimes b)\neq\psi_{p,W_2}(f\otimes b) $. This shows that $W$ can not be ergodic.\\
1) $\Rightarrow$ 2). Assume that $W$ is extremal and that there are a stationary ergodic $p\in \mathcal{P}(A^{\zz},\Sigma_c )$, $a\in (0,1)$ and two stationary states $\psi_1, \psi_2$ on  $B(A^{\zz},\Sigma_c)\otimes \mathcal{B}^{\zz} $ with $\psi_1\neq \psi_2$ and
\begin{equation}\label{conv-decomp-2}
  \psi_{p,W}=a\psi_1 + (1-a)\psi_2.
\end{equation}
Our goal is to construct stationary cq-channels $W_1 , W_2 : A^{\zz}\times\mathcal{B}^{\zz}\to\cc $, $W_1\neq W_2$,  with
\begin{equation}\label{goal-1}
W=a W_1 + (1-a)W_2,
\end{equation}
and 
\begin{equation}\label{goal-2}
 \psi_i(f\otimes b)=\int f(x)W_i (x,b)p(dx),\quad i=1,2
\end{equation}
for $f\in B(A^{\zz},\Sigma_c), b\in \mathcal{B}^{\zz}$. This would contradict the assumed extremality of $W$ and we are done. To this end, by ergodicity of $p$ and using the convex decomposition of $\psi_{p,W}$ in eq. (\ref{conv-decomp-2}) we see that
\begin{equation}\label{equal-p}
\psi_i\upharpoonright B(A^{\zz}, \Sigma_c)=p, \quad i=1,2, 
\end{equation}
holds.\\
Fix $b\in \mathcal{B}^{\zz}$ and define a linear functional $l_{b,i}: B(A^{\zz},\Sigma_c )\to\cc$ by
\[ l_{b,i}(f):= \psi_{i}(f\otimes b).\]
It is obvious that
\[||l_{b,i}||\le ||b||,\]
holds. $b$ can be written as a complex linear combination of four positive elements $b_k\in\mathcal{B}^{\zz}$, $k=1,\ldots ,4$, e.g. consider first $b=\frac{1}{2}(b+b^{\ast})+i\frac{1}{2i}(b-b^{\ast})$ and then apply functional calculus to both hermitian summands to obtain positive and negative parts of them (cf. \cite{kadison-1} section 4.1). For each $f\ge 0$ we have then
\begin{eqnarray}\label{abs-cont}
l_{b_k,i}(f)&=&\psi_i (f\otimes b_k)\le ||b_k|| \psi_{i}(f\otimes \idn_{\mathcal{B}^{\zz}})\nonumber\\
& =& ||b_k||p(f)\quad (\textrm{by eq. } (\ref{equal-p}) ).
\end{eqnarray}
Thus by the dominated convergence theorem, for each sequence $f_j\searrow 0$, $f_j\in B(A^{\zz},\Sigma_c)$, we have $l_{b_k ,i}(f_ j)\searrow 0$ and this implies that the functional $l_{b_k,i}$ is representable by a unique finite measure $p_{b_k ,i}$ on $(A^{\zz},\Sigma_c)$ and this finite measure is absolutely continuous with respect to $p$ by eq. (\ref{abs-cont}). Therefore $l_{b,i}$ is representable by a complex measure $p_{b,i}$ which is absolutely continuous with respect to $p$ (cf. \cite{rudin} for definitions and background information). 
Thus we can infer from the Radon-Nikodym Theorem (cf. \cite{rudin} for a version concerning complex measures) that there is a $\cc$-valued Borel measurable function $\rho_{i}(\cdot , b)$ with
\begin{equation}\label{radon-nikodym}
  l_{b,i}(f)=\psi_{i}(f\otimes b) =\int f(x)\rho_{i}(x,b)p(dx).
\end{equation}
In the next step we show that $\rho_i$ can be changed on a set of $p-$measure $0$ such that the resulting map $W_i$ has the following properties:
\begin{enumerate}
\item[a)] $W_{i}(x, \cdot)$ is a state on $\mathcal{B}^{\zz}$. 
\item[b)] $W_{i}(T_{in}x,b)=W_{i}(x,T_{out}b)$ for all $x\in A^{\zz}$, $b\in\mathcal{B}^{\zz}$.
\end{enumerate}
We consider a basis $(u_{ij})_{i,j=1}^{d}$ of $\mathcal{B}$ consisting of matrix units (i.e. $u_{ij}u_{kl}=\delta_{j,k}u_{il}$, $u_{ij}^{\ast}=u_{ji}$ and $\sum_{i=1}^{d}u_{ii}=\idn_{\mathcal{B}}$) and, for $n\in\nn$, we consider the tensor product basis $(u_{i_{-n}^{n}j_{-n}^{n}})$ of $\mathcal{B}^{[-n,n]}$ built up from the single site basis above. Let $V_{n}$ denote the set of all complex linear combinations of these basis vectors in $\mathcal{B}^{[-n,n]} $ whose real and imaginary parts are rational numbers. Thus $V_n$ is a \emph{countable} linear space over the field $\mathbb{Q}+i\mathbb{Q}$. If $n'\ge n$ there is natural inclusion $V_n\subset V_{n'}$ which is induced by the quasi-local structure of $\mathcal{B}^{\zz}$. Then the linear space over $\mathbb{Q}+i\mathbb{Q}$ given by $V:=\bigcup_{n\in\nn}V_n$ is also countable and dense in $\mathcal{B}^{\zz}$. The idea is to construct $W_i$ first on $V$ and then to show that the extension to $\mathcal{B}^{\zz}$ inherits the properties a), b) above. This parallels the construction of the regular conditional probability in probability theory, the difference being only that we use an algebraic language.\\
An inspection of the integral formula, eq. (\ref{radon-nikodym}), together with some standard measure-theoretic arguments (cf. Theorem 1.40 in \cite{rudin}), show that for $s,t\in \mathbb{Q}+i\mathbb{Q}$, $b,b'\in V$ and $b''\in V$ with $\idn\ge b''\ge 0$ the sets
\[L_{s,t,b,b'}:=\{x\in A^{\zz}: \rho_i(x, sb+tb')\neq s\rho_i (x,b)+t\rho_i (x,b')\}, \]
\[P_{b''}:=\{x\in A^{\zz}:\rho_i (x,b'')\notin [0,+\infty) \} \]
and
\[ U:=\{ x\in A^{\zz}:\rho_i (x,\idn)\neq 1 \} \]
have $p-$measure $0$. Since $V$ is countable, union of all these sets has $p-$measure $0$. Thus for
\begin{equation}\label{zero-set-1}
N_1:=U \cup \bigcup_{s,t,b,b'}L_{s,t,b,b'}\cup \bigcup_{b''\in V:\idn \ge b''\ge 0}P_{b''}, 
\end{equation}
we have $p(N_1)=0$.\\
For $b\in V$ define
\[ S(b):=\{x\in A^{\zz}: \rho_i (T_{in}x, b)\neq \rho_i (x, T_{out}b) \}. \]
Using $\psi_i (T_{in}f\otimes T_{out}b)=\psi_{i}(f\otimes b)$, eq. (\ref{radon-nikodym}), the change of variable formula and $T_{in}-$invariance of $p$ it is easily seen that
\[ \rho_i (x,b)=\rho_i (T_{in}^{-1}x, T_{out}b)\quad p-\textrm{a.s.}, \]
and this is equivalent to
\[ \rho_i (T_{in}x, b)= \rho_i (x, T_{out}b)\quad p-\textrm{a.s.} \]
since our shifts are invertible. We have $p(N_2)=0$ for $N_2:=\bigcup_{b\in V}S(b)$. 
Set
\[ E(b):= \{x\in A^{\zz}: W (x, b)\neq a\rho_{1}(x,b)+(1-a)\rho_{2} (x, b) \},\]
for $b\in V$.\\
Eq. (\ref{conv-decomp-2}) and eq. (\ref{radon-nikodym}) imply that $p(E(b))=0$ for all $b\in V$. Set $N_3:=\bigcup_{b\in V}E(b)$. Then $p(N_3)=0$ holds.\\
For $N_1$ given by eq. (\ref{zero-set-1}), $N_2$, and $N_3$ set $N_4:=N_1\cup N_2\cup N_3$ and define $N:=\bigcup_{k\in\zz}T_{in}^{k}N_4$. Then $N_4\subset N$ and $T_{in}N= N$, moreover we have $p(N)=0$ since $p$ is $T_{in}-$invariant. We are now in position to define $W_i$ on $V$:
\begin{eqnarray}\label{def-wi}
 W_{i}(x,b) =\left\{ \begin{array}{ll}
\rho_{i}(x,b) & \textrm{if } x\in N^{c}\\
W(x,b) & \textrm{else}
\end{array}\right. ,
\end{eqnarray}
for $i=1,2$ and $b\in V$. $W_i (x,\cdot)$ is by construction a positive linear normalized functional on $V$, and $W_{i}(T_{in}x,b)=W_{i}(x, T_{out}b)$ for all $b\in V$ and all $x\in A^{\zz}$. Note that the completion of each $V_n$ is $\mathcal{B}^{[-n,n]}$ ($V_n$ is even a normed $^{\ast}-$subalgebra of $\mathcal{B}^{[-n,n]}$ since we have used a basis consisting of matrix units in construction of $V_n$). It is fairly standard fact that we can extend each $W_{i}(x,\cdot)$ to $\mathcal{B}^{[-n,n]}$ while preserving its norm, i.e. $ ||W_{i}(x,\cdot )||=1$ on each $\mathcal{B}^{[-n,n]}$. This in turn gives us a linear bounded extension from $\mathcal{B}^{\textrm{loc}}=\bigcup_{n\in\nn}\mathcal{B}^{[-n,n]}$ to $\mathcal{B}^{\zz}$ with $||W(x,\cdot )||=1$. But then we can apply theorem 4.3.2 from \cite{kadison-1}, which states that each bounded linear functional $l$ defined on a self-adjoint subspace containing $\idn$ of a $C^{\ast}-$algebra with $||l||=l(\idn)$ is positive. Thus we have constructed two stationary cq-channels $W_1,W_2:A^{\zz}\times \mathcal{B}^{\zz}\to \cc$ with properties (\ref{goal-1}), (\ref{goal-2}) and $W_1\neq W_2$ since by our hypothesis we have $\psi_1\neq\psi_2$. This concludes our proof.  \end{proof}
Let $\mathcal{E}$ denote the convex set of all completely positive unital maps $E:B(A^{\zz},\Sigma_c)\otimes\mathcal{B}^{\zz}\to B(A^{\zz},\Sigma_c) $ which satisfy eq. (\ref{factorization}) and $E\circ (T_{in}\otimes T_{out})=T_{in}\circ E$, while $\mathcal{K}$ stands for the convex set of all completely positive unital maps $K:\mathcal{B}^{\zz}\to B(A^{\zz},\Sigma_c)$ with $K\circ T_{out}=T_{in}\circ K$. Then note that the extremality of $W$ is brought forward to the associated completely positive maps $K$ and $E$ via Theorem \ref{affine-corresp-of-all-things}. Thus we have following
\begin{corollary}\label{ergodicity-of-K-and-E}
Let $W:A^{\zz}\times \mathcal{B}^{\zz}\to \cc$ be a stationary cq-channel, and consider the associated completely positive unital maps $K:\mathcal{B}^{\zz}\to B(A^{\zz},\Sigma_c)$ and $E:B(A^{\zz},\Sigma_c)\otimes\mathcal{B}^{\zz}\to B(A^{\zz},\Sigma_c)  $. Then following statements are equivalent:
\begin{enumerate}
\item $W$ is ergodic.
\item $W$ is extremal in the convex set of stationary cq-channels.
\item $K$ is extremal in $\mathcal{K}$.
\item $E$ is extremal in $\mathcal{E}$.
\end{enumerate}
\end{corollary}
In light of our discussion preceding Theorem \ref{heisenberg} we see that, if we replace the input algebra $B(A^{\zz},\Sigma_c)$ by an arbitrary quasi-local algebra $\mathcal{A}^{\zz}$, the ergodicity of a quantum channel $K:\mathcal{B}^{\zz}\to\mathcal{A}^{\zz}$ should be defined in the following way; $K$ is stationary, i.e. $K\circ T_{out}=T_{in}\circ K$, and $K$ is an extreme point in the convex set of stationary quantum channels. The equality of channels is defined in analogy to eq. (\ref{channel-equality-2}), i.e. channels $K_1,K_2:\mathcal{B}^{\zz}\to\mathcal{A}^{\zz}$ are said to be equal if for all $a\in\mathcal{A}^{\zz}$, all $b\in\mathcal{B}^{\zz}$ and each stationary state $\varphi$ on $\mathcal{A}^{\zz}$ 
\begin{equation}\label{channel-equality-quantum}
  \varphi (aK_1(b))=\varphi (aK_2(b))
\end{equation}
holds.\\
Before closing this subsection we give an example that emphasizes the important role played by equality definition (\ref{channel-equality-2}) for cq-channel $K$, or equivalently for the maps $W$ and $E$.\\
\emph{Example.} Let us consider the commutative $C^{\ast}-$algebra $B(A^{\zz},\Sigma_c)$ with $A=\{0,1\}$. Let $a\in A^{\zz}$ be the periodic sequence given by $a:=(\ldots,0,1,0,1,\ldots )$ and consider its shifted version $Ta$, where $T:A^{\zz}\to A^{\zz}$ denotes the usual (left) shift on doubly-infinite sequences. Denoting by $\delta_{a}$ the point measure concentrated on the sequence $a$ we define $q:=\frac{1}{2}(\delta_{a}+\delta_{Ta})$. It is easily seen that $q$ is stationary ergodic with respect to $T$. Let us consider the channel $K:B(A^{\zz},\Sigma_{c})\to B(A^{\zz},\Sigma_c)$ with $K(f):=q(f)\idn$ where $\idn$ denotes the identity in $B(A^{\zz},\Sigma_c)$. It is then obvious that for all $f,g\in B(A^{\zz},\Sigma_c)$ and all probability distributions $p$ on $A^{\zz}$ we have $p(gK(f))=p(g)q(f)$, i.e. the joint input-output state is a product state. If we choose $p=q$ then it is easily verified that the set $I_1:=\{ (a,Ta), (Ta,a))$ fulfills $(T\otimes T)^{-1}I_1=I_1$ and $(q\otimes q)(I_1)=\frac{1}{2}$, thus $q\otimes q$ is not ergodic. Hence the channel $K$ is not ergodic. However, it is clear that $K$ maps stationary ergodic measures to stationary ergodic measures, namely to $q$ which is stationary  ergodic.\\
If we define the equality of channels by the requirement that for all $f\in B(A^{\zz}, \Sigma_c)$ and all stationary probability measures $p\in\mathcal{P}(A^{\zz},\Sigma_c)$
\[ p(K_1(f))=p(K_2(f))\]
holds, then it is obvious that the channel $K$ from above is extreme point in the set of stationary channels with respect to this notion of equality.
On the other hand, if we use the notion of equality from eq. (\ref{channel-equality-quantum}) then it is readily checked that for
\[  W_{1}(x,g) =\left\{ \begin{array}{ll}
1_{\{a \}}(x)g(Ta)+1_{\{Ta \}}(x)g(a) & \textrm{if } x\in \{a,Ta\}\\
q(g) & \textrm{else}
\end{array}\right.  \]
and
\[ W_{2}(x,g) =\left\{ \begin{array}{ll}
1_{\{a \}}(x)g(a)+1_{\{Ta \}}(x)g(Ta) & \textrm{if } x\in \{a,Ta\}\\
q(g) & \textrm{else}
\end{array}\right. \]
we have a convex decomposition of the channel $K$ into stationary channels with the weights $(\frac{1}{2},\frac{1}{2})$ in the sense of definition in (\ref{channel-equality-quantum}) with $W_1\neq W_2$.
\subsection{Continuity Properties of CQ-Channels}\label{continuity}
The idea that continuity properties of channels play a crucial role in establishing coding theorems is well known in information theory and goes back to the classic paper by McMillan \cite{mcmillan}. Subsequent development of this idea in classical information theory showed that the most fruitful notion of continuity is that with respect to the $\bar{d}-$distance as was demonstrated by Gray and Ornstein \cite{gray-ornstein}. At the present time we do not have a notion of distance for quantum states having similarly nice properties as the $\bar{d}-$distance in the classical setting. E.g. the $\bar{d}-$distance is extremely sensitive to ergodic/mixing properties of probability measures and is much weaker than variational distance. Moreover the entropy rates are $\bar{d}-$continuous. Nice introductions to this notion of distance and its application in ergodic/information theory can be found in the monographs \cite{shields} by Shields and \cite{gray} by Gray. In this paper we will restrict ourselves to the variational distance, which can be extended to quantum states without causing any problems (cf. \cite{gray-ornstein} for corresponding classical definitions of continuity of channels with respect to the variational distance and disadvantages of them).  \\ 
A cq-channel $W$ is called \emph{causal} if for each $n\in\zz $, $b\in \mathcal{B}^{(-\infty,n]}$ and all $x,\tilde x\in A^{\zz} $ with $x_{i}=\tilde x_{i}$ for $i\le n$
\[W(x,b)=W(\tilde x ,b),\]
holds. $W$ is called \emph{input memoryless} if for each $n\in\zz$, $b\in \mathcal{B}^{[n,\infty)}$ and all $x,\tilde x\in A^{\zz} $ with $x_{i}=\tilde x_{i}$ for $i\ge n$ the channel fulfills
\[W(x,b)=W(\tilde x ,b).\]
\begin{remark}\label{rem-continuity-channel}
We see at this point immediately that the dependence of the channel on past and future inputs is intimately connected with the continuity properties of the map $W$. E.g. if the channel $W$ is input memoryless and causal (IMC in what follows) then for each $b\in \mathcal{B}^{[n,n+k]}$ and each $x\in A^{\zz}$ the function $W(x,b)$ depends merely on the coordinates $x_n, x_{n+1}, \ldots x_{n+k}$, and thus can be identified with a continuous function on $A^{\zz}$. In view of Lemma \ref{trivial-1} and the continuity of the linear map $K:\mathcal{B}^{\zz}\to B(A^{\zz},\Sigma_{c}) $ (which is ensured by its positivity) it is immediately clear that $K( \mathcal{B}^{\zz} )\subset C(A^{\zz})$, the continuous $\cc$-valued functions equipped with $||\cdot||_{\infty}$-norm. Consequently, in this case the map $A^{\zz}\ni x\mapsto W(x,b)$ is continuous for any fixed $b\in \mathcal{B}^{\zz}$.
\end{remark} 
Although the requirement that the channel $W$ acts causally seems to be natural, the restriction to input memoryless channels would be a serious limitation in many situations. However, we expect that the effect of the input letters far in the past should not affect much present and future outputs. This motivates our next definition.\\
Channel $W$ is said to have \emph{decaying input memory} (DIM channel for short) if for each $\eps>0$ there is an integer $m (\eps)$ such that for all $n\in \zz$ and all $b\in \mathcal{B}^{[n,\infty)}$ with $0\le b\le\idn$
\begin{equation}\label{decay-memory}
  |W(x,b)-W(x',b)|\le \eps,
\end{equation}
 whenever $x_i=x'_{i}$ for $i\ge n-m$ for all $m\ge m(\eps)$. This can be compactly restated as
 \begin{equation}\label{decay-memory-alter}
   \lim_{m\to\infty}\sup_{x,y: x_{n-m}^{\infty}=y_{n-m}^{\infty}}d_{v,n}(W_x,W_y)=0,
 \end{equation}
for all $n\in\zz$ where $W_{x}:= W(x,\cdot )$ and 
\begin{equation}\label{var-distance}
  d_{v,n}(W_x,W_y):=\sup_{b\in \mathcal{B}^{[n,\infty)}:0\le b\le \idn}|W(x,b)-W(y,b)|,
\end{equation}
is the usual variational distance for quantum states. Note that the set maximization in eq. (\ref{var-distance}) is performed over can be replaced by the set of orthogonal projections in $\mathcal{B}^{[n,\infty)} $ since the functional which is maximized is convex and projections are extreme points of the set $\{b\in \mathcal{B}^{[n,\infty)}:0\le b\le \idn\} $ (see \cite{davies} Lemma 2.3).\\
A channel $W$ has \emph{decaying input memory and anticipation} (DIMA) if for each $\eps>0$ there are non-negative integers $m (\eps), a(\eps)$ such that for all $n,\in \zz, k\in\nn$ and all $b\in \mathcal{B}^{[n,n+k]}$ with $0\le b\le\idn$ eq. (\ref{decay-memory}) holds whenever $x_i=x'_{i} $ for $n-m\le i\le n+k+a $ and all $m\ge m(\eps)$, $a\ge a(\eps)$. Again, this can be equivalently described by
\begin{equation}\label{dima-def}
 \lim_{m,a\to\infty}\sup_{x,y:x_{n-m}^{n+k+a}=y_{n-m}^{n+k+a} }d_{v,n,k}(W_x , W_y)=0, 
\end{equation}
for all $n\in \zz, k\in\nn$ where $d_{v,n,k}(\cdot , \cdot)$ is defined analogously to $d_{v,n}$ in eq. (\ref{var-distance}) the difference being only that we replace $ \mathcal{B}^{[n,\infty)} $ by $\mathcal{B}^{[n,n+k]} $.\\
Our next lemma shows that the associated map $K$ from Lemma \ref{trivial-1} of each causal DIM or DIMA channel $W$ maps into the set $C(A^{\zz})$.
\begin{lemma}\label{trivial-2}
Let $W:A^{\zz}\times \mathcal{B}^{\zz}\to\cc$ be a cq-channel and let $K:\mathcal{B}^{\zz}\to B(A^{\zz},\Sigma_{c}) $ be the associated map from Lemma \ref{trivial-1}. 
\begin{itemize}
\item[a)] Following assertions are equivalent:
\begin{enumerate}
\item $K(\mathcal{B}^{\zz})\subseteq C(A^{\zz})$.
\item For each $b\in \mathcal{B}^{\zz}$
\[ \lim_{n\to\infty}\sup_{x,y\in A^{\zz}: x_{-n}^{n}=y_{-n}^{n}}|W(x,b)-W(y,b)|=0. \]
\end{enumerate}
\item[b)] If $W$ is DIMA then $K(\mathcal{B}^{\zz})\subseteq C(A^{\zz}) $
\end{itemize}
\end{lemma}
\begin{proof} a) Simple consequence of the fact that cylinder sets are open.\\
b) That $K(b)\in C(A^{\zz})$ for $b\in \mathcal{B}^{[n,n+k]}$ holds is obvious since every such $b$ can be written as $b=b_1+ib_2$ with hermitian $b_1 , b_2 \in \mathcal{B}^{[n,n+k]} $ and each hermitian element of $\mathcal{B}^{[n,n+k]} $ can be written as a linear combination of orthogonal projections in $\mathcal{B}^{[n,n+k]} $ (spectral theorem) so that our DIMA condition, eq. (\ref{dima-def}), applies. For a general $b\in\mathcal{B}^{\zz}$, by definition of the quasi-local algebra $\mathcal{B}^{\zz}$, there are sequences $n_i \in \zz$, $k_i\in \nn$ and $b_i\in \mathcal{B}^{[n_i , n_i +k_i]} $ with $\lim_{i\to\infty}||b_i-b||=0$. By continuity of $K$ this implies that $\lim_{i\to\infty}||K(b_i)-K(b)||_{\infty}=0$ and thus that $K(b)\in C(A^{\zz})$. \end{proof}
\emph{General Hypotheses}
\begin{enumerate}
\item Since we will be concerned only with DIMA cq-channels we will henceforth consider only $C(A^{\zz})\subsetneq B(A^{\zz},\Sigma_c)$ and will consider the joint input-output state over $C(A^{\zz})\otimes \mathcal{B}^{\zz}$. The ergodicity will be always defined with respect to this algebra.
\item All channels are assumed to be stationary
\item It is easily inferred from the results in \cite{paulsen}, chapter 12, and \cite{kadison-ringrose}, sections 11.3 and 11.4 that $C(A^{\zz})\otimes \mathcal{B}^{\zz}$ is $^{\ast}-$isomorphic to the quasi-local algebra $(\mathfrak{F}(A)\otimes \mathcal{B})^{\zz}$. This is basically due to the facts that $C(A^{\zz})\otimes \mathcal{B}^{\zz} $ can be seen as the inductive limit of $C^{\ast}-$algebras $(\mathfrak{F}(A)^{[-n,n]}\otimes \mathcal{B}^{[-n,n]})_{n\in\nn}$ and each of these algebras is $^{\ast}-$isomorphic to $(\mathfrak{F}(A)\otimes \mathcal{B})^{[-n,n]}$. We shall therefore consider $C(A^{\zz})\otimes \mathcal{B}^{\zz}$ w.l.o.g. as a quasi-local algebra in what follows and the results from \cite{bksss} apply to this situation.
\end{enumerate}
Consider an input memoryless causal (IMC) cq-channel $K: \mathcal{B}^{\zz}\to C(A^{\zz})$. We see immediately that in this situation the cq-channel can be described by a family of maps $W^{n}:A^n\times \mathcal{B}^{n}\to \cc$, $n\in\nn$, such that $W^{n}(x^{n},\ \cdot \ )$ is a state on $\mathcal{B}^{n}$ for each $x^{n}:=(x_{1},\ldots ,x_{n})\in A^{n}$. Now, suppose that $W$ is ergodic. If the input measure $p$ is ergodic and if we denote its marginal distributions on $A^n$ by $p^{n}$ then we obtain for the marginal input-output state
\[\psi^{n}_{p,W}(1_{C}\otimes b)=\sum_{x^n\in C}p^{n}(x^n)W^{n}(x^n,b),\]
and these marginal input-output states $\{\psi^{n}_{p,W}\}_{n\in\nn}$ define an ergodic state on $C(A^{\zz})\otimes \mathcal{B}^{\zz}$ which we also denote by $\psi_{p,W}$ for notational simplicity. It is easily seen that the state $\psi^{n}_{p,W}$ corresponds to the following density operator
\begin{equation}\label{imc-dens-matrix}
D_{p,W}^{n}=\sum_{x^n\in A^{n}}p^{n}(x^n)|x^n\rangle\langle x^n|\otimes D_{x^n},
\end{equation}
where $D_{x^n}$ denotes the density operator of the state $W^{n}(x^n,\ \cdot \ )$, $x^n=(x_1,\ldots ,x_n)\in A^{n}$ and $|x^n\rangle =e_{x_{1}}\otimes e_{x_{1}}\otimes \cdots \otimes e_{x_{n}} $ for some fixed orthonormal basis $\{e_{i}\}_{i=1}^{|A|}$ of $\cc^{|A|}$.\\
A \emph{code} $\mathcal{C}=(u_{i},b_{i})_{i=1}^{M}$ consists of sequences $u_{1},\ldots , u_{M}\in A^{n}$ (code words) and a family of positive semi-definite operators $b_1,\ldots ,b_{M}\in \mathcal{B}^{n}$ (decoding operators) with $\sum_{i=1}^{M}b_{i}\le\idn$.\\
The \emph{error probability} of a code $\mathcal{C}=(u_{i},b_{i})_{i=1}^{M} $ is given by
\[e(\mathcal{C}):=\max_{i=1,\ldots ,M}\sup_{x\in [u_i]}(1-W(x, b_i)), \]
where $[u_i] $ denotes the cylinder set generated by the sequence $u_i$. Average error probability is given by
 \begin{equation}\label{average-error}
\bar{e}(\mathcal{C}):=\frac{1}{M}\sum_{i=1}^{M}\sup_{x\in [u_i]}(1-W(x, b_i)).
\end{equation}
A real number $R$ is said to be an \emph{achievable rate} for the cq-channel $W$ if there is a sequence of codes $(\mathcal{C}_{n})_{n\in\nn}$ with
\[\liminf_{n\to\infty}\frac{1}{n}\log M_n\ge R,\]
and
\[\lim_{n\to\infty}e(\mathcal{C}_{n})=0.\]
The \emph{(weak) capacity} $C(W)$ of the cq-channel $W$ is defined as the least upper bound of achievable rates.\\
The expression for the error probabilities of an IMC cq-channels simplifies to
\[c(\mathcal{C})=\max_{i=1,\ldots ,M}(1-W^n(u_i,b_i)). \]
\subsection{Examples}\label{examples}
\emph{Example 1.} Our first example is discrete memoryless cq-channel as considered by \cite{holevo-2, schumacher, winter}. Clearly, such channels are input memoryless, causal and stationary ergodic.\\
\emph{Example 2.} Another interesting class of channels are those with Markovian correlated noise from \cite{macchiavello, bowen-mancini, datta-dorlas}.  Let $(E_y)_{y\in I} $ be a finite family of completely positive unital maps $E_y:\mathcal{B}\to \mathcal{A}$ where $\mathcal{B},\mathcal{A} $ denote algebras of linear operators over suitable finite-dimensional Hilbert spaces. Moreover let us consider a stationary irreducible aperiodic Markovian probability measure $\mu \in\mathcal{P}(I^{\nn},\Sigma_c)$ with stationary distribution $q$ and transition matrix $Q=q(\cdot |\cdot )$. For each $n\in \zz$ and $k\in \nn$ we define a completely positive unital map $E_{n,k}:\mathcal{B}^{[n,n+k-1]}\to \mathcal{A}^{[n,n+k-1]} $ by
\[ E_{n,k}:=\sum_{y^k\in I^k}\mu^k (y^k)E_{y_1}\otimes \cdots \otimes E_{y_k}.\]
This family determines a unique completely positive unital map $E:\mathcal{B}^{\zz}\to\mathcal{A}^{\zz}$ whose restriction to $\mathcal{B}^{[n,n+k-1]} $ coincides with $E_{n,k}$. It is clear by definition of $E$ that $E\circ T_{\mathcal{B}}=T_{\mathcal{A}}\circ E$ holds, where $T_{\mathcal{B}}$ resp. $T_{\mathcal{A}} $ denote the shifts on $\mathcal{B}^{\zz} $ resp $\mathcal{A}^{\zz} $, i.e. the quantum channel $E$ is stationary. Furthermore, the condition $E\upharpoonright \mathcal{B}^{[n,n+k-1]}=E_{n,k} $ means that the channel is input memoryless and causal. Consider the dual map $E':\mathcal{S}(\mathcal{A}^{\zz})\to \mathcal{S}(\mathcal{B}^{\zz}) $ which is defined by
\[ (E'(\varphi))(b):=\varphi(E(b))\quad (b\in \mathcal{B}^{\zz},\varphi\in \mathcal{S}(\mathcal{A}^{\zz}) ),\]
and local dual maps $E'_{n,k}:\mathcal{S}(\mathcal{A}^{[n,n+k-1]})\to \mathcal{S}(\mathcal{B}^{[n,n+k-1]}) $ which are given by
\[E'_{n,k} =\sum_{y^k\in I^k}\mu^k (y^k)E'_{y_1}\otimes \cdots \otimes E'_{y_k},  \]
where $E'_{y}$ is defined by
\[ \textrm{tr}(E'_{y}(D)b):=\textrm{tr}(DE_{y}(b)),  \]
for all $b\in\mathcal{B}$ and all density operators $D$ in $\mathcal{A}$.\\
For a given finite family $(D'_{a})_{a\in A}$ of density operators in $\mathcal{A}$ let us consider the states $\varphi_{x}$ on $\mathcal{A}^{\zz}$, $x\in A^{\zz}$, whose restrictions to $\mathcal{A}^{[n,n+k-1]}$ have density operators $D'_{x_{n}}\otimes \cdots \otimes D'_{x_{n+k-1}}$. Set
\[ W(x,b):=E'(\varphi_{x})(b)\quad (b\in \mathcal{B}^{\zz} ).\]
It is clear that for $b\in \mathcal{B}^{[n,n+k-1]}$
\[ W(x,b)=\sum_{y^k\in I^k}\mu^k (y^k)\textrm{tr}((E'_{y_1}(D'_{x_1})\otimes \cdots\otimes  E'_{y_k}(D'_{x_k}))b)  \]
holds, and thus the cq-channel $W$ is input memoryless, causal and stationary.\\
Due to our assumption that the Markovian measure is irreducible and aperiodic we know that $\lim_{n\to\infty}(Q^n)_{y',y}= q(y)$ exponentially fast for all $y,y'\in I$. Using this fact it is easily shown that for all $b_1,b_2\in \mathcal{B}^{[n,n+k]}$ and all $x\in A^{\zz}$
 \[\lim_{l\to\infty}|W(x,b_1 T_{out}^{l}b_2)-W(x,b_1)W(x,T_{out}^{l}b_2)|=0,\]
exponentially fast. By approximating $b_1,b_2\in\mathcal{B}^{\zz}$ by local observables it is immediately clear that
 \[\lim_{l\to\infty}|W(x,b_1 T_{out}^{l}b_2)-W(x,b_1)W(x,T_{out}^{l}b_2)|=0,\]
i.e. the channel $W$ is output mixing in classical terminology. Due to this fact and mimicking the classical calculation (see \cite{adler, gray} Lemma 9.4.3) using definition of the joint state, eq. (\ref{joint-state}), it is easily shown that $W$ is ergodic. \\
Note that a similar construction starting with an arbitrary stationary \emph{mixing} probability measure $\mu\in \mathcal{P}(I^{\nn},\Sigma_c) $ leads to a stationary ergodic IMC cq-channel. Hence our Theorems \ref{coding-th-direct} and  \ref{weak-converse} apply to this whole class of cq-channels.\\
\emph{Example 3.} We can easily modify last example to obtain channels with finite input memory; We just have to use an irreducible aperiodic stationary Markovian measure of order $k$ in our construction.\\
Further examples can be found in \cite{kretschmann-werner} and the references therein.
\section{Results on Typical Projections}\label{preliminary}
In this section we give some auxiliary results that are repeatedly used in the rest of the paper.\\
We start with a non-commutative version of the fact that the intersection of two sets that have high probability must also be highly likely. We use the Hilbert-Schmidt inner product on linear operators that is given by $\langle a,b\rangle_{HS}:=\textrm{tr}(a^{\ast}b)$ for $a,b\in \mathcal{L}(\hr)$. Moreover we need the notion of partial trace for density operators acting on $\hr_1\otimes \hr_2$ which is defined as follows: Let $D\in \mathcal{L}(\hr_1\otimes \hr_2)\simeq \mathcal{L}(\hr_1)\otimes \mathcal{L}(\hr_2)$ be a density operator. The partial traces of $D$ are uniquely determined density operators $D_i$, $i=1,2$ in $\mathcal{L}(\hr_i)$ which fulfill
\[ \textrm{tr}_{\hr_1}(D_1a_1)=\textrm{tr}_{\hr_1\otimes \hr_2}(D(a_1\otimes \idn_{\hr_2}))\]
and
\[\textrm{tr}_{\hr_2}(D_2a_2)=\textrm{tr}_{\hr_1\otimes \hr_2}(D (\idn_{\hr_1}\otimes a_2)) \]
 for all $a_1\in\mathcal{L}(\hr_1)$ resp. $a_2\in \mathcal{L}(\hr_2)$. Usual notation for $D_1$ is $\textrm{tr}_{2}(D)$ and similarly for $D_2$.   
\begin{lemma}\label{gentle-pinching}
1) Let $D\in\mathcal{L}(\hr)$  be an operator with $0\le D\le \idn$ and $\textrm{tr}(D)\le 1$ and let $q_1,q_2\in \mathcal{L}(\hr)$ be projections with $\textrm{tr}(Dq_{i})=1-\eps_{i}$, $i=1,2$. Then
\[ \textrm{tr}(Dq_2q_1q_2)\ge 1-\eps_{1} -2\sqrt{\eps_{2}}.\]
2) Let $D\in \mathcal{L}(\hr_{1}\otimes \hr_{2})$ be a density operator and let $D_{1}:=\textrm{tr}_{2}(D)$ resp. $D_{2}:=\textrm{tr}_{1}(D)$ be the reduced density operators. Then for any projections $q_{i}\in\mathcal{L}(\hr_{i})$, $i=1,2$, with $\textrm{tr}(D_{i}q_{i})=1-\eps_{i}$ we have
\[\textrm{tr}(D(q_1\otimes q_2))\ge 1-\eps_{1}-\sqrt{\eps_{2}}.\]
\end{lemma}
\begin{proof} 1) The proof consists of an elementary application of the Cauchy-Schwarz inequality for the Hilbert-Schmidt inner product. Indeed, note that $\idn=q_2+(\idn-q_2)$, then we have
  \begin{eqnarray}\label{gentle-pinching-1}
    1-\eps_1&=& \textrm{tr}(Dq_1)=\textrm{tr}(D\idn q_1)\nonumber \\
            &=& \textrm{tr}(D(q_2+(\idn-q_2))q_1)\nonumber \\
            &\le& |\textrm{tr}(Dq_2q_1)|+|\textrm{tr}(D(\idn-q_2)q_1)  |.   
  \end{eqnarray}
Note that 
\[|\textrm{tr}(D(\idn-q_2)q_1)  |=|\textrm{tr}(D^{1/2}(\idn-q_2)q_1D^{1/2} )\le \sqrt{\eps_2},\] 
where we have applied the Cauchy-Schwarz inequality to $a^{\ast}:=D^{1/2}(\idn-q_2) $ and $b:=q_1D^{1/2} $ and we have used that $\textrm{tr}(D(\idn-q_2))=\eps_2$ and $\textrm{tr}(Dq_1)\le 1$ hold. Thus the inequality (\ref{gentle-pinching-1}) above can be rewritten as
\[ 1-\eps_1-\sqrt{\eps_2}\le |\textrm{tr}(Dq_2q_1 )| .  \]
Using $\idn=q_2+(\idn-q_2)$ again we obtain from this inequality
\begin{eqnarray*}
 1-\eps_1-\sqrt{\eps_2}&\le & |\textrm{tr}(Dq_2q_1 )| \\
                       &\le& |\textrm{tr}(Dq_2q_1q_2)|+|\textrm{tr}(Dq_2q_1(\idn-q_2))|\\
                       &=& |\textrm{tr}(Dq_2q_1q_2)|\\
                       & &+|\textrm{tr}(D^{1/2}q_2q_1(\idn-q_2)D^{1/2})|.
\end{eqnarray*}
The last term can be upper bounded by $\sqrt{\eps_2}$ as is easily seen from $q_2q_1q_2\le\idn$ and the Cauchy-Schwarz inequality applied to $a^{\ast}:= D^{1/2}q_2q_1 $ and $b:= (\idn-q_2)D^{1/2}$. We are done now because 
\[ |\textrm{tr}(Dq_2q_1q_2)|= \textrm{tr}(Dq_2q_1q_2),   \]
which in turn follows from $ q_2q_1q_2\ge 0$.\\
2) Note that by our assumption and by the definition of the partial trace we have
\begin{eqnarray*}
  1-\eps_{1}&=&\textrm{tr}(D_{1}q_1)=\textrm{tr}(D(q_1\otimes \idn))\\
&=& \textrm{tr}(D(q_1\otimes q_2))+ \textrm{tr}(D(q_1\otimes (\idn-q_2)))\\
&=& \textrm{tr}(D(q_1\otimes q_2))+ \textrm{tr}(D(q_1\otimes \idn)(\idn\otimes (\idn-q_2)))
\end{eqnarray*}
The desired inequality is then a consequence of $\textrm{tr}(D_{2}q_2)=1-\eps_{2} $, the definition of the partial trace and the Cauchy-Schwarz inequality for the Hilbert-Schmidt inner product.\end{proof}
Our next lemma states that the restriction of the local states to a sequence of high-probability subspaces does not affect the v. Neumann entropy rate. The notions of states, stationarity and v. Neumann entropy rate are introduced in the appendix, especially in section \ref{quasi-local-app}.
\begin{lemma}\label{entropy-pinching}
Let $\psi\in \mathcal{S}(\mathcal{A}^{\zz})$ be a stationary state with v. Neumann entropy rate $s$ where $\mathcal{A}^{\zz} $ is a quasi-local algebra constructed from a finite dimensional $C^{\ast}$-algebra $\mathcal{A}$. Let $(q_{n})_{n\in \nn}$, $q_{n}\in\mathcal{A}^{n}$, be a sequence of projections with $\lim_{n\to\infty}\psi(q_{n})=1$. Then we have
\[ \lim_{n\to\infty}\frac{1}{n}S(q_{n}D_{\psi^{n}}q_{n})=s.\]
\end{lemma}
\begin{proof} Define $D_{n}:=q_n D_{\psi^{n}}q_n+ q_{n}^{\perp}D_{\psi^{n}}q_{n}^{\perp}$, where $q_{n}^{\perp} $ denotes the projection onto the orthogonal complement of the range of $q_n$. Then we know that (cf. \cite{lindblad})
\begin{equation}\label{entropy-increase}
 S(D_{\psi^{n}})\le S(D_n)= S(q_n D_{\psi^{n}}q_n )+S(q_{n}^{\perp}D_{\psi^{n}}q_{n}^{\perp} ),
\end{equation}
holds. Now, using
\[ S\left( \frac{q_{n}^{\perp}D_{\psi^{n}}q_{n}^{\perp}}{\textrm{tr}(D_{\psi^{n}}q_{n}^{\perp})}\right)\le \log\textrm{tr}(q_{n}^{\perp} ),  \]
 and $\lim_{n\to\infty}\psi(q_{n}^{\perp})=0 $ it is easily seen that
\begin{equation}\label{entropy-qperp-0}
 \lim_{n\to\infty}\frac{1}{n}S(q_{n}^{\perp}D_{\psi^{n}}q_{n}^{\perp})=0. 
\end{equation}
On the other hand, consider a spectral decomposition $D_{\psi^{n}}=\sum_{i=1}^{d^{n}}\lambda_{i}e_{i}$ with one-dimensional mutually orthogonal projections $e_i$. Recall that the entropy is almost convex, i.e. for any probability vector $a:=(a_{1},\ldots ,a_{k})$ and any set of density operators $D_1,\ldots ,D_k$ we have $S(\sum_{i=1}^{k}a_iD_i)\le\sum_{i=1}^{k}a_iS(D_i)+H(a) $, (cf. \cite{landford}) where $H(a)$ denotes the Shannon entropy of the probability vector $a$. Inserting the spectral decomposition above in (\ref{entropy-increase}) and using almost convexity we arrive at
\begin{eqnarray*}
  0\le S(D_{n})-S(D_{\psi^{n}})&\le & \sum_{i=1}^{d^{n}}\lambda_{i}S(q_n e_{i}q_n+ q_{n}^{\perp}e_{i}q_{n}^{\perp} )\\
& &+H((\lambda_{1},\ldots ,\lambda_{d^{n}}))-S(D_{\psi^{n}})\\
&=& \sum_{i=1}^{d^{n}}\lambda_{i}S(q_n e_{i}q_n+ q_{n}^{\perp}e_{i}q_{n}^{\perp} )\\
&\le & \log (2)=1,
\end{eqnarray*}
where we have used $H((\lambda_{1},\ldots ,\lambda_{d^{n}}))=S(D_{\psi^{n}}) $ in the third line, and the observation that the dimension of the range of each $q_n e_{i}q_n+ q_{n}^{\perp}e_{i}q_{n}^{\perp} $ is at most $2$ which implies the last inequality. This shows that
\[ \lim_{n\to\infty}\frac{1}{n}S(D_{n})=s,\]
holds. Combining this with (\ref{entropy-qperp-0}) and right hand half of (\ref{entropy-increase}) leads to the desired conclusion of the lemma.\end{proof}
Our main ingredient for the proof of the coding theorem will be the quantum version of the Shannon-McMillan theorem (a.k.a. quantum AEP) for ergodic states on quasi-local algebras. We refer to \cite{bksss} for a proof.
\begin{theorem}[Quantum AEP]\label{SM} 
Let \(\psi\) be a stationary ergodic state with entropy rate $s$ on  quasi-local algebra \( {\cal A}^{\zz}\) built up from a finite dimensional $C^{*}$-algebra $\mathcal{A}$. Then
for any \( \eps > 0\) there is a sequence of orthogonal projections $(t_{n,\eps})_{n\in\nn}$, $t_{n,\eps}\in
{\cal A}^{n}$, such that for all sufficiently large $n$ hold:
\begin{enumerate}
\item $\psi^{n}(t_{n,\eps})\ge 1-\eps $,
\item for each one-dimensional orthogonal projection $q\in {\cal A}^{n} $ which is dominated by $t_{n,\eps}$, i.e. $q\le t_{n,\eps}$ we have 
  \begin{displaymath}
    2^{-n(s+\eps)}<\psi^{n}(q)<2^{-n(s-\eps)},
  \end{displaymath}
\item $ 2^{n(s-\eps)}< \textrm{tr}(t_{n,\eps})<2^{n(s+\eps)}  $.
\end{enumerate}
Moreover, the \emph{entropy typical subspace} given by the range of each $t_{n,\eps} $ can be spanned by those eigenvectors of $D_{\psi^{n}}$ associated with the eigenvalues $\mu_{i,n}$ of $D_{\psi^{n}}$ which satisfy $2^{-n(s+\eps)}<\mu_{i,n}<2^{-n(s-\eps)} $.$\quad \Box$
\end{theorem}
An alternative, equivalent version of this theorem can be stated  for dimension covering exponents (cf. \cite{bksss} for proof), which are defined in the following way: Consider any state $\psi$ on \( {\cal A}^{\zz}\) and $\eps\in (0,1)$. The dimension covering exponents are given by
\[\beta_{\eps,n}(\psi):=\min\{\log\textrm{tr}(q): q\in\mathcal{A}^{n} \textrm{ projection, } \psi^{n}(q)\ge 1-\eps\}.\]
\begin{proposition}\label{dim-covering-sm}
Let $\psi$ be a stationary ergodic state on ${\cal A}^{\zz} $ with v. Neumann entropy rate $s$. Then for any $\eps\in (0,1)$
\[ \lim_{n\to\infty}\frac{1}{n} \beta_{\eps,n}(\psi)=s\]
holds.
\end{proposition}
For the following lemma we need some preliminary notation. Let $\mathcal{A}^{\zz}$ denote a quasi-local algebra built up from finite dimensional $C^{\ast}$-algebra $\mathcal{A}$. We consider a state $\psi$ on $\mathcal{A}^{\zz} $ and any sequence $(q_{n})_{n\in\nn}$ of orthogonal projections with $q_n\in\mathcal{A}^{n}$. To the state $\psi$ and sequence $(q_n)_{n\in\nn}$ we associate a family of sub-normalized states $q_{n}D_{\psi^{n}}q_{n}$ and for each $n\in\nn$ we consider a diagonalization of these sub normalized states:
\begin{eqnarray}\label{restrict-diag}
q_{n}D_{\psi^{n}}q_{n}=\sum_{i=1}^{\textrm{tr}(q_n)}\lambda_{i,n}q_{i,n},
\end{eqnarray}
where $\lambda_{i,n}$ resp. $q_{i,n}$ denote eigenvalues resp. one-dimensional projections onto an orthonormal basis of range of $q_n$ consisting of eigenvectors of $q_{n}D_{\psi^{n}}q_{n}$. We abbreviate $I_n:=\{1,\ldots ,\textrm{tr}(q_n)\}$.
\begin{lemma}\label{high-prob-vs-sm}
Let $\psi$ be a stationary ergodic state on $\mathcal{A}^{\zz}$ with v. Neumann entropy rate $s$ and let $(q_{n})_{n\in\nn} $ be a sequence of projections with $q_n\in\mathcal{A}^{n} $ and $\lim_{n\to\infty}\psi(q_n)=1$. Then for each $\eps>0$ there is sequence of projections $(r_n(\eps))_{n\in\nn}$ with $r_n(\eps)\le q_n$ with
\begin{enumerate}
\item $\lim_{n\to\infty}\psi(r_n(\eps))=1$.
\item For any one-dimensional projection $r\le r_n(\eps)$ we have
\[ 2^{-n(s+\eps)}< \textrm{tr}(q_{n}D_{\psi^{n}}q_{n}r  )=\psi^{n}(r)<2^{-n(s-\eps)}.\]
\end{enumerate}
The range of $r_n(\eps)$ is spanned by those eigenvectors of $q_{n}D_{\psi^{n}}q_{n}$ which satisfy
\[  2^{-n(s+\eps)}< \lambda_{i,n}<2^{-n(s-\eps)}.\]
\end{lemma}
\begin{proof} The proof is quite similar to the proof of Lemma 3.3 in \cite{bksss} and is, in fact, much easier in the present situation because we can use now the full Shannon-McMillan theorem for ergodic quantum states that was not at our disposal in \cite{bksss} (the Shannon-McMillan theorem for ergodic quantum states was even proved there). For readers convenience we give main steps of this proof. First note that
\begin{eqnarray}\label{eq:high-prob-1}
 S(q_{n}D_{\psi^{n}}q_{n})=-\sum_{i\in I_n}\lambda_{i,n}\log \lambda_{i,n},
\end{eqnarray}
and that according to Lemma \ref{entropy-pinching} 
\begin{eqnarray}\label{eq:high-prob-2}
\lim_{n\to\infty}\frac{1}{n}S(q_{n}D_{\psi^{n}}q_{n})=s.
\end{eqnarray} 
For $\delta>0$ we consider the set
\[ B_{1,n,\delta}:=\left\{i\in I_n: \lambda_{i,n}\ge 2^{-n(s-\delta)}\right\},\]
and the projection
\[r_{1,n,\delta}:=\sum_{i\in B_{1,n,\delta}}q_{i,n}.\]
Then it is clear that
\begin{eqnarray}\label{eq:high-prob-3}
\frac{1}{n}\log\textrm{tr}(r_{1,n,\delta})\le s-\delta \qquad \forall \ n\in\nn,
\end{eqnarray}
and $\psi^{n}(r_{1,n,\delta})=\textrm{tr}(q_{n}D_{\psi^{n}}q_{n}r_{1,n,\delta} )$ hold.
If we had $\limsup_{n\to \infty}\psi^{n}(r_{1,n,\delta_{0}})=a>0$ for some $\delta_{0}>0$ then eq. (\ref{eq:high-prob-3}) would contradict Proposition \ref{dim-covering-sm}.
Hence we must have
\begin{eqnarray}\label{eq:high-prob-4}
\lim_{n\to\infty}\psi^{n}(r_{1,n,\delta})=0 \qquad \forall \ \delta >0.  
\end{eqnarray}
For $\sigma , \delta >0$ we define the following sets
\[B_{2,n,\sigma}:=\left\{i\in I_n: \lambda_{i,n}\le 2^{-n(s+\sigma)}\right\},\]
and
\begin{equation*}
T_{n,\delta,\sigma}:= \left\{i\in I_n: 2^{-n(s+\sigma)}< \lambda_{i,n}
 <  2^{-n(s-\delta)}\right\}. 
\end{equation*}
Then it is obvious that for $i\in B_{2,n,\sigma}$ resp. $i\in T_{n,\delta,\sigma} $ we have
\[ -\frac{1}{n}\lambda_{i,n}\log\lambda_{i,n}\ge (s+\sigma)\lambda_{i,n}\]
 respectively 
\[-\frac{1}{n}\lambda_{i,n}\log\lambda_{i,n}\ge (s-\delta)\lambda_{i,n}.\]
Combining these inequalities with eq. (\ref{eq:high-prob-1}) and (\ref{eq:high-prob-4}) we are led to
\begin{eqnarray}\label{eq:high-prob-5}
\frac{1}{n}S(q_{n}D_{\psi^{n}}q_{n})&\ge & (s+\sigma)\psi^{n}(r_{2,n,\sigma}) \nonumber\\
&+ & (s-\delta)\psi^{n}(r_{n,\sigma,\delta})+o(1),
\end{eqnarray}
where
\[r_{2,n,\sigma}:=\sum_{i\in B_{2,n,\sigma}}q_{i,n},\textrm{ and } r_{n,\sigma,\delta}:= \sum_{i\in T_{n,\delta,\sigma} }q_{i,n}.\]
If we had $\limsup_{n\to\infty}\psi (r_{2,n,\sigma_{0}})=a>0$ for some $\sigma_{0}>0$ then after taking limit $n\to\infty$  along a suitable subsequence in eq. (\ref{eq:high-prob-5}) we would have
\[s\ge s+ \sigma_{0}a - \delta (1-a)>s, \]
for all sufficiently small $\delta>0$, a contradiction. Thus, we have
\[\lim_{n\to\infty}\psi^{n}(r_{n,\sigma,\delta})=1\qquad \forall \ \sigma,\delta >0. \]
Now, if we define $r_{n}(\eps):=r_{n,\eps,\eps}$ we have the desired sequence of projections. The second item of lemma is easily verified since the range of $r_{n}(\eps)$ is spanned by eigenvectors of $q_{n}D_{\psi^{n}}q_{n}  $ the eigenvalues of which satisfy $2^{-n(s+\eps)}< \lambda_{i,n}
 <  2^{-n(s-\eps)} $. \end{proof}

\section{Coding Theorem for IMC Channels}\label{coding-imc}
Our goal in this section is to show that for each stationary ergodic IMC cq-channel $W$ we have
\begin{equation}\label{cap-limit}
C(W)=C_{\textrm{Holevo}}(W),
\end{equation}
where
\begin{equation}\label{holevo-cap-def}
C_{\textrm{Holevo}}(W):= \lim_{n\to\infty}\frac{1}{n}C_{n}(W) = \sup_{n}\frac{1}{n}C_{n}(W), 
\end{equation}
and
\[ C_{n}(W):=\max_{p^n\in\mathcal{P}(A^{n})}\chi (p^n,W^n)\]
with the well known Holevo information
\[\chi (p^n, W^n):=S(D_{q(p)}^{n})-\sum_{x^n\in A^n}p^{n}(x^n)S(D_{x^n}), \]
and $D_{q(p)}^{n}:=\sum_{x^n\in A^n}p^{n}(x^n)D_{x^n} $. An equivalent formula for the Holevo information can be given as follows:
\[\chi (p^n, W^n)=S(D_p^n)+S(D_{q(p)}^n)-S(D_{p,W}^n),  \]
where $D_p^n=\sum_{x^n\in A^n}p^n(x^n)|x^n\rangle \langle x^n|$ and $D_{p,W}^n$ is given in eq. (\ref{imc-dens-matrix}). The equality in the last formula holds since $S(D_{p,W}^{n} )=S(D_p^n)+\sum_{x^n\in A^n}p^{n}(x^n)S(D_{x^n}) $ (cf. \cite{nielsen-chuang} Th. 11.8).
Note that the limit in  eq. (\ref{holevo-cap-def}) exists and is actually equal to $\sup_{n}\frac{1}{n}C_{n}(W)  $ since we have
\[ C_{n+m}(W)\ge C_{n}(W)+C_{m}(W)\]
which is easily deduced from the fact that
\begin{equation}\label{superadditivity}
 \chi (p^n\otimes p'^m, W^{n+m})\ge \chi (p^n, W^n)+\chi (p'^m, W^m),
\end{equation}
which in turn is a consequence of the subadditivity of the formally defined conditional v. Neumann entropy (cf. \cite{nielsen-chuang} Theorem 11.16) and our assumption that the cq-channel is stationary and IMC. \\
\begin{remark}\label{subadditivity-con-entropy}
 For a probability distribution $p$ on $A^n$ the formal conditional v. Neumann Entropy is, in our case, given by
\[S(p|q(p)):= S(D_{p,W^n}^{n})-S(D_{q(p)}^n),\]
with
\[D_{p,W^n}^{n}=\sum_{x^n\in A^n}p(x^n)|x^n\rangle \langle x^n|\otimes D_{x^n},  \]
and
\[D_{q(p)}^n= \sum_{x^n\in A^n}p(x^n)D_{x^n}.  \]
The subadditivity of this quantity together with stationarity and IMC property of the channel means in the present situation that for $p\in \mathcal{P}(A^n)$, $p'\in \mathcal{P}(A^m)$
\[ S(p\otimes p'|q(p\otimes p' ))\le S(p|q(p))+S(p'|q(p'))\]
holds.
\end{remark}
Before entering the proof of the direct part of the coding theorem we need some further preliminary lemmas. Let us consider a stationary IMC cq-channel $W$ and a stationary probability measure $p$ on $(A^{\zz},\Sigma_{c})$. Then since the states $\psi_{p,W}, p$ and $\psi_{q(p)}:=\psi_{p,W}\upharpoonright \mathcal{B}^{\zz}$ are stationary we know that their v. Neumann entropy rates exist. Let us denote by $D_{p,W}^{n}, D_{p}^{n}$ resp. $D_{q(p)}^{n}$ the density operators of these states when restricted to the algebras over $n$-blocks.\\
Let us define formally the information rate
\begin{eqnarray}\label{information-rate} 
i(p,W)&:=& \lim_{n\to\infty}\frac{1}{n}\chi (p^n, W^n)\nonumber \\
&=&\lim_{n\to\infty}\frac{1}{n}(S(D_p^n)+S(D_{q(p)}^n)-S(D_{p,W}^{n} ) )\nonumber \\
&=& s(\psi_{p})+s(\psi_{q(p)})-s(\psi_{p,W}). 
\end{eqnarray}
Note that $s(\psi_{p})=h(p)$ where $h(p)$ denotes the Shannon entropy rate of the probability measure $p$. Due to the results in \cite{robinson} the entropy rates in eq. (\ref{information-rate}) exist even for a larger class of periodic states. Furthermore let us introduce
\[ C_{\textrm{per}}(W):=\sup_{p \textrm{ periodic}}i(p,W),\]
\[ C_{\textrm{stat}}(W):=\sup_{p \textrm{ stationary}}i(p,W),\]
and 
 \begin{equation}\label{def-c-erg}
 C_{\textrm{erg}}(W):=\sup_{p \textrm{ stationary ergodic}}i(p,W).
\end{equation}
\begin{lemma}\label{equality-of-capacities}
Let $W:A^{\zz}\times \mathcal{B}^{\zz}\to\cc$ be a stationary IMC cq-channel. Then we have
\[C_{\textrm{Holevo}}(W)=  C_{\textrm{per}}(W) =C_{\textrm{stat}}(W)= C_{\textrm{erg}}(W).\]
\end{lemma}
\begin{proof} It is clear that 
\[C_{\textrm{Holevo}}(W)\ge  C_{\textrm{per}}(W)  \ge C_{\textrm{stat}}(W)\ge C_{\textrm{erg}}(W),\]
holds. Thus we only need to prove $C_{\textrm{Holevo}}(W)=C_{\textrm{erg}}(W) $. By definition of the Holevo capacity eq. (\ref{holevo-cap-def}) we can find for any $\delta >0$ a probability distribution $p_{\delta}\in\mathcal{P}(A^t)$ with
\begin{equation}\label{first}
C_{\textrm{Holevo}}(W)-\delta\le \frac{1}{t}\chi (p_{\delta},W^t).
\end{equation} 
Set $p'= p_{\delta}^{\otimes\infty}$, then the probability measure $p'$ is $t$-periodic.
Now we define
\[ p:=\frac{1}{t}\sum_{i=0}^{t-1}p'\circ T_{in}^{-i}.\]
It is easily seen by standard arguments that $p$ is stationary ergodic. In what follows we use the abbreviation $p_i:=p'\circ T_{in}^{-i} $. For each $i\in \{0,1,\ldots ,t-1\}$ and for each $n\in\nn$ the distributions $p_i^n$ can be written as
\[p_i^n=p_{l_{i}}\otimes p_{\delta}^{\otimes k_{i}}\otimes p_{r_{i}}, \]
where $n=l_{i}+k_i t +r_i$ with $0\le l_i + r_i<2t$. Note that $k_i$ depends on $n$ and that $\lim_{n\to\infty}\frac{k_i}{n}=\frac{1}{t}$ holds for all $i\in \{0,\ldots ,t-1\}$.\\
Using concavity of the Holevo information with respect to the input distribution and eq. (\ref{superadditivity}) we obtain following chain of inequalities:
\begin{eqnarray*}
  \frac{1}{n}\chi (p^n, W^n)&=&  \frac{1}{n}\chi (\frac{1}{t}\sum_{i=0}^{t-1}p_i^n,W^n)\\
&\ge & \frac{1}{n}\frac{1}{t}\sum_{i=0}^{t-1}\chi(p_i^n,W^t )\\
&=& \frac{1}{n}\frac{1}{t}\sum_{i=0}^{t-1}\chi(p_{l_{i}}\otimes p_{\delta}^{\otimes k_{i}}\otimes p_{r_{i}}, W^{l_i+k_i t+r_i})\\
&\ge & \frac{1}{n}\frac{1}{t}\sum_{i=0}^{t-1}( \chi (p_{l_i},W^{l_i})+ \chi (p_{r_i},W^{r_i})\\
& & + k_i\chi(p_{\delta}, W^{t}))\\
&\ge & \frac{\min_{i\in\{0,\ldots ,t-1\}}k_i}{n}\chi(p_{\delta}, W^{t})+o(1).
\end{eqnarray*}
This yields
\[ i(p,W)=\lim_{n\to\infty}\frac{1}{n}\chi (p^n, W^n)\ge \frac{1}{t}\chi(p_{\delta}, W^{t}).\]
Combining this with eq. (\ref{first}) and definition of $C_{\textrm{erg}}(W)$, eq. (\ref{def-c-erg}), we obtain
\[C_{\textrm{erg}}(W)\ge  C_{\textrm{Holevo}}(W)-\delta .\]
Since the left side of this inequality does not depend on $\delta$ and since $\delta>0$ was arbitrary we can conclude that
\[C_{\textrm{erg}}(W)\ge  C_{\textrm{Holevo}}(W).\]\end{proof}
We will need some simple properties of projections in $\mathfrak{F}(A)\otimes \mathcal{B}$ in the proof of the next lemma, where $\mathfrak{F}(A)$ denotes the set of $\cc -$valued functions on $A$. We again identify $\mathfrak{F}(A)$ with $\cc^{|A|}$. It is elementary to show that each operator $a\in \mathfrak{F}(A)\otimes \mathcal{B}$ can be written as
\[ a=\sum_{x\in A}|x\rangle\langle x|\otimes a_x,\]
for appropriate $a_x\in \mathcal{B}$. Similarly, $a^{\ast}=a$ iff $a_x^{\ast}=a_x$ for all $x\in A$. Moreover it holds that $a^2=a$ iff $a_x^{2}=a_x$ for all $x\in A$. Thus we can conclude that each projection $t \in \mathfrak{F}(A)\otimes \mathcal{B}$ has a unique representation
\begin{equation}\label{no-entanglement}
t=\sum_{x\in A}|x\rangle\langle x|\otimes t_x,
\end{equation}
with projections $t_x\in\mathcal{B}$. This is an analogon to the representation of a set in a Cartesian product of finite sets as a union of its sections.
\begin{lemma}[Probability bounds]\label{prob-bounds}
Let $W:A^{\zz}\times \mathcal{B}^{\zz}\to \cc$ be a stationary ergodic IMC cq-channel and let $p\in \mathcal{P}(A^{\zz},\Sigma_{c})$ be a stationary ergodic probability measure. Then for each $\eps >0$ there is a sequence of orthogonal projections $(j_{n}(\eps))_{n\in\nn}$, $j_{n}(\eps)\in \mathfrak{F}(A^n)\otimes \mathcal{B}^{n}$, with
\begin{displaymath}
  \lim_{n\to\infty}\psi_{p,W}(j_n (\eps))=1,
\end{displaymath}
and which possesses following additional properties: For each $n\in\nn$ there is subset $T_n\subseteq A^n$ and a set of orthogonal projections $(c_{x^n})_{x^n\in T_n}$ in $\mathcal{B}^n$ with $j_n (\eps)=\sum_{x^n\in T_n}|x^n\rangle \langle x^n|\otimes c_{x^n}$ and
\begin{enumerate}
\item $\lim_{n\to\infty}p^n(T_n)=1$.
\item For each $x^n\in T_n$ we have
\[2^{-n(s(\psi_{p})+\eps_n )}<p^n(x^n)<2^{-n(s(\psi_{p})-\eps_n )},\]
with $s(\psi_p)=h(p)$ and an appropriate sequence with $1>\eps_n\searrow 0$.
\item $W^n(x^n, c_{x^n})=1-\delta_{n}$ for all $x^n\in T_n$ and a suitable sequence with $\lim_{n\to\infty}\delta_{n}= 0$.
\item For each $x^n\in T_n$,all sufficiently large $n\in\nn$, and each one-dimensional projection $e\le c_{x^n}$ we have
\begin{equation}\label{eq:prob-bounds}
2^{-n(s(q(p)|p)+\eps)}\le W^n(x^n, e)\le 2^{-n(s(q(p)|p)-\eps)}, 
\end{equation}
with $s(q(p)|p):=s(\psi_{p,W})-s(\psi_p) $. Consequently, the dimension of the range of $c_{x^n} $ can be bounded by
\[2^{n(s(q(p)|p)-\eps +\frac{1}{n}\log (1-\delta_{n}))}\le \textrm{tr}(c_{x^n})\le  2^{n(s(q(p)|p)+\eps)}.  \]
\end{enumerate}
\end{lemma}
\begin{proof} Choose an appropriate sequence $1\ge \eps_n \searrow 0$ such that the quantum AEP, Theorem \ref{SM}, holds simultaneously for $\psi_{p,W}$, $\psi_{p}$ and $\psi_{q(p)}$ with $\eps_n$ instead of $\eps$ and denote by $t_n$, $t_{p,n}$ resp. $t_{q,n}$ the resulting entropy typical projections of $\psi_{p,W}$, $\psi_{p}$ resp. $\psi_{q(p)}$. An application of Lemma \ref{gentle-pinching}.2 yields that
\[\psi_{p,W}^{n}(t_{p,n}\otimes t_{q,n})\ge 1-\eps_n-\sqrt{\eps_n},\]
and Lemma \ref{gentle-pinching}.1 then implies that
\begin{equation}\label{j-typ-1}
\psi_{p,W}^{n}((t_{p,n}\otimes t_{q,n})t_n(t_{p,n}\otimes t_{q,n}) )= 1-\eta_{n},  
\end{equation}
with $\eta_{n}\le \eps_{n}+2\sqrt{\eps_n +\sqrt{\eps_n} }$. Set
\[ t'_n:=R((t_{p,n}\otimes t_{q,n})t_n(t_{p,n}\otimes t_{q,n})  ),\]
where $R(a)$ denotes the projection onto the range of the hermitian operator $a$. Then we have
\begin{equation}\label{j-typ-2}
t'_n \le t_{p,n}\otimes t_{q,n}, 
\end{equation}
\begin{equation}\label{j-typ-3}
  \textrm{tr}(t'_n)\le \textrm{tr}(t_{p,n}\otimes t_{q,n}  ), \textrm{ tr}(t_n),
\end{equation}
and
\begin{equation}\label{j-typ-4}
  \psi_{p,W}^n(t'_n)\ge 1-\eta_n ,
\end{equation}
by eq. (\ref{j-typ-1}) and eq. (\ref{j-typ-2}). Lemma \ref{high-prob-vs-sm} gives us for each $\eps>0$ a sequence of projections $(t''_n)_{n\in\nn}$ satisfying $t''_n \le t'_n $, $t''_n\in \mathfrak{F}(A^n)\otimes \mathcal{B}^n$, and with
\begin{equation}\label{j-typ-5}
 \psi_{p,W}^{n}(t''_n)\ge 1-\eta'_n \quad \lim_{n\to\infty}\eta'_n = 0,
\end{equation}
and 
\begin{equation}\label{j-typ-entropy}
 2^{-n(s(\psi_{p,W})+\frac{\eps}{2})}<\psi_{p,W}^{n}(r)<2^{-n(s(\psi_{p,W})-\frac{\eps}{2})} 
\end{equation}
for any one-dimensional orthogonal projection $r\le t''_n$. We assume w.l.o.g. that $\eta'_n>0$ for all $n\in \nn$.\\
It is readily seen from (\ref{no-entanglement}) that for each $x^n\in A^n$ we have
\begin{equation}\label{j-typ-6}
 (|x^n\rangle\langle x^n|\otimes \idn) t''_n= t''_n (|x^n\rangle\langle x^n|\otimes \idn),
\end{equation}
where $\idn$ denotes the identity in $\mathcal{B}^n$. This yields
\begin{equation}\label{j-typ-7}
  t''_n=\sum_{x^n\in A^n}(|x^n\rangle\langle x^n|\otimes \idn) t''_n (|x^n\rangle\langle x^n|\otimes \idn),
\end{equation}
and if we define projections $c_{x^n}$ by 
\begin{equation}\label{j-typ-8}
 |x^n\rangle\langle x^n|\otimes c_{x^n}= (|x^n\rangle\langle x^n|\otimes \idn) t''_n (|x^n\rangle\langle x^n|\otimes \idn),
\end{equation}
we can write
\begin{equation}\label{j-typ-9}
 t''_n= \sum_{x^n\in P_n}|x^n\rangle\langle x^n|\otimes c_{x^n}.
\end{equation}
with
\begin{equation}\label{p-def}
P_n:=\{x^n\in A^n: c_{x^n}\neq 0\}.
\end{equation}
For the set $T_n$ given by
\begin{equation}\label{j-typ-10}
  T_n:=\{ x^n\in P_n: W^n(x^n,c_{x^n})\ge 1-\sqrt{\eta'_n}\},
\end{equation}
we see that
\begin{equation}\label{j-typ-12}
  p^n(T_n^c)\le \sqrt{\eta'_n},
\end{equation}
holds, since by eq. (\ref{j-typ-5}) and eq. (\ref{j-typ-9}) we have
\begin{eqnarray*}
 1-\eta'_n &\le & \psi_{p,W}^{n}(t''_n)=\sum_{x^n\in A^n}p^n(x^n)W^n(x^n, c_{x^n})\\
&\le & p^n(T_n^c)(1-\sqrt{\eta'_n} )+ p^n(T_n^c)\\
&=& 1- \sqrt{\eta'_n}p^n(T_n^c).
\end{eqnarray*}
Set
\[ r_n:=\sum_{x^n\in T_n}|x^n\rangle \langle x^n|,\]
and
\begin{equation}\label{j-def}
j_n (\eps):=(r_n\otimes \idn)t''_n (r_n\otimes \idn)=\sum_{x^n\in T_n}|x^n\rangle \langle x^n|\otimes c_{x^n} .
\end{equation}
Applying Lemma \ref{gentle-pinching} one more time we arrive at
\[\lim_{n\to\infty}\psi_{p,W}^n(j_n (\eps))=1.\]
By our construction we have $t''_n\le t'_n\le t_{p,n}\otimes t_{q,n}$ (see eq. (\ref{j-typ-2})). Since $r_n\otimes \idn$ commutes with $t''_n$, i.e. $(r_n\otimes \idn)t''_n=t''_n(r_n\otimes \idn)$, we have 
\[ (r_n\otimes \idn)t''_n(r_n\otimes \idn)\le t''_n \le t_{p,n}\otimes t_{q,n} ,\]
and therefore by the right hand side of eq. (\ref{j-def}) for each $x^n\in T_n\subseteq P_n$
\[|x^n\rangle \langle x^n|\otimes c_{x^n}\le \sum_{x^n\in T_n}|x^n\rangle \langle x^n|\otimes c_{x^n}\le t_{p,n}\otimes t_{q,n}. \]
This yields
\[r_n\le t_{p,n},\]
since $T_n\subseteq P_n$ (cf. eq. (\ref{p-def})), consequently, for each one-dimensional projection $|x^n\rangle\langle x^n|\le r_n$
\begin{equation}\label{j-typ-13}
 2^{-n(s(\psi_p)+\eps_n)}< \psi_{p}(|x^n\rangle\langle x^n|  )=p^n(x^n)< 2^{-n(s(\psi_p)-\eps_n)}.
\end{equation}
Additionally eq. (\ref{j-typ-10}) and eq. (\ref{j-typ-12}) yield that 
\[ \lim_{n\to\infty}\psi^n(r_n)=\lim_{n\to\infty}p^n(T_n)=1\]
hold.
Note that $j_n (\eps)\le t''_n$ and for each one-dimensional projection $e\in \mathcal{B}^n$ we have
\[p^n(x^n)W^n(x^n,e)=\psi_{n,W}^n(|x^n\rangle\langle x^n|\otimes e).\]
Thus by eq. (\ref{j-typ-entropy}) and eq. (\ref{j-typ-13}) for each $x^n\in T_n$ and for each one-dimensional projection $e\le c_{x^n}$ we obtain eq. (\ref{eq:prob-bounds}) for all sufficiently large $n$ for which $\eps_n < \frac{\eps}{2}$ holds.\end{proof}
\begin{theorem}[Coding theorem: direct part]\label{coding-th-direct}
Let $W: A^{\zz}\times \mathcal{B}^{\zz}\to\cc$ be a stationary ergodic IMC cq-channel. Then there is a sequence of codes $(\mathcal{C}_{n})_{n\in\nn}$ with
\begin{equation}\label{c-achiev}
  \liminf_{n\to\infty}\frac{1}{n}\log M_{n}\ge C_{\textrm{Holevo}}(W) ,
\end{equation}
with $\lim_{n\to\infty}e(\mathcal{C}_{n})=0$. I.e. $C_{\textrm{Holevo}}(W) $ is an achievable rate for the cq-channel $W$ and consequently we have $C(W)\ge C_{\textrm{Holevo}}(W) $.
\end{theorem}
\begin{proof} The proof is virtually the same as Winter's proof \cite{winter} for the direct part of the coding theorem for memoryless cq-channel. The difference to the present situation is only that we use our Lemma \ref{prob-bounds} instead of frequency typical and frequency conditionally typical subspaces in Winter's setting which are defined in analogy to frequency typical and frequency conditionally typical subsets in the approach of Wolfowitz (see \cite{wolfowitz}).\\
For readers convenience and since we shall need this argument in section \ref{coding-dima} we give the main steps.\\
From Lemma \ref{equality-of-capacities} we know that
\[C_{\textrm{Holevo}}(W)= C_{\textrm{erg}}(W).  \] 
Thus, for each $\delta >0$ we can find a stationary ergodic probability measure $p\in\mathcal{P}(A^{\zz},\Sigma_{c})$ with
\[ i(p,W)\ge C_{\textrm{erg}}(W)-\frac{\delta}{2}.\]
It suffices to show that for each $\delta>0$ and $\lambda\in (0,1)$ there is a sequence of codes $(\mathcal{C}_n)_{n\in\nn}$ with $M_n$ code words such that
\begin{enumerate}
\item $e(\mathcal{C}_n)\le\lambda$ and
\item $M_n\ge 2^{n(C_{\textrm{erg}}(W)-\delta )}$
\end{enumerate}
for all $n\ge n(\delta,\lambda)$ with an appropriately chosen $n(\delta,\lambda)\in\nn $.\\
With the notation from Lemma \ref{prob-bounds} let $n$ be large enough to ensure that $\delta_n <\frac{\lambda}{2}$.\\
According to Lemma \ref{prob-bounds}, with $\eps=\frac{\delta}{4}$, there is $x^n\in T_n$ and $c_{x^n}\in\mathcal{B}^n$ with
\[W^n(x^n, c_{x^n})=\textrm{tr}(D_{x^n}c_{x^n})\ge 1-\lambda.\]
Set $u_1:=x^n$ and $b_1:=c_{x^n}$. Note that $b_1$ is a projection. \\
In the next step choose $x^n\in T_n$ and $c_{x^n}$ with
\[ W^n(x^n,(\idn-b_1)c_{x^n}(\idn-b_1))\ge 1-\lambda,\]
and set $u_2:=x^n$ and $b_2:=R((\idn-b_1)c_{x^n}(\idn-b_1))$ (projection onto the range of $(\idn-b_1)c_{x^n}(\idn-b_1) $). Then it is clear that
\[ W^n(u_2, b_2)\ge 1-\lambda \]
since $b_2\ge (\idn-b_1)c_{u_2}(\idn-b_1) $.\\
If $u_1 , \ldots , u_k$ and $b_1 ,\ldots ,b_k$ are already constructed then choose $x^n\in T_n$ with
\[ W^n(x^n,(\idn-\sum_{i=1}^{k}b_i)c_{x^n} (\idn-\sum_{i=1}^{k}b_i))\ge 1-\lambda,\]
and set 
\[u_{k+1}:=x^n \textrm{ and } b_{k+1}:=R((\idn-\sum_{i=1}^{k}b_i)c_{x^n} (\idn-\sum_{i=1}^{k}b_i)).\]
Continue this procedure until no further prolongation of the code is possible. Note that each $b_i$ is an orthogonal projection and that $b_i b_j=\delta_{i,j}b_i$ holds.\\
Let us write $\mathcal{C}_n=(u_i,b_i)_{i=1}^{M_n}$ for the resulting code and set
\[b_n:=\sum_{i=1}^{M_n}b_i.\]
For $\eta:=\min\{1-\lambda, \frac{\lambda^2}{16}\}>0$ we claim that
\begin{equation}\label{shadow}
  W^n(x^n,b_n)=\textrm{tr}(D_{x^n}b_n)\ge \eta \quad \forall x^n\in T_n.
\end{equation}
This is clear for code words. If we had 
\[ \textrm{tr}(D_{x^n}(\idn-b_n))\ge 1-\eta \]
for some $x^n\in T_n\setminus \{u_1,\ldots ,u_{M_n}\}$, then we had 
\begin{eqnarray*}
 \textrm{tr}(D_{x^n}(\idn-b_n)c_{x^n}(\idn-b_n)  )&\ge& 1-\delta_{n}-2\sqrt{\eta}\\
&\ge &1-\frac{\lambda}{2}-2\sqrt{\frac{\lambda^2}{16}}\\
&=& 1-\lambda
\end{eqnarray*} 
by Lemma \ref{gentle-pinching} and our restriction to those $n$ for which $\delta_n<\frac{\lambda}{2} $ holds. The last inequality implies that we could prolong our code, what is not possible by our code construction. Averaging eq. (\ref{shadow}) with $p^n$ we obtain
\[ \psi_{q(p)}^n(b_n)=\sum_{x^n\in A^n}p^n(x^n)W^n(x^n,b_n)\ge p^n(T_n)\eta>\frac{\eta}{2},\]
for all sufficiently large $n$. Then Proposition \ref{dim-covering-sm} yields that
\begin{equation}\label{dir-part-1}
  \textrm{tr}(b_n)\ge 2^{n(s(\psi_{q(p)} )-\frac{\delta}{4})}
\end{equation}
for all $n$ which are large enough.\\
On the other hand, we see by our code construction that $\textrm{tr}(b_i)\le\textrm{tr}(c_{u_i})$ for all $i\in \{1, \ldots , M_n\}$ and it follows from Lemma \ref{prob-bounds}.4 that for all sufficiently large $n$
\begin{equation}\label{dir-part-2}
  \textrm{tr}(b_n)\le M_n 2^{n(s(q(p)|p)+\frac{\delta}{4} )}
\end{equation}
holds. Combining eq. (\ref{dir-part-1}) and eq. (\ref{dir-part-2}) we obtain
\[M_n\ge 2^{n(i(p,W)-\frac{\delta}{2})}\ge 2^{n(C_{\textrm{erg}}(W)-\delta )} . \]
This concludes our proof since $C_{\textrm{Holevo}}(W)= C_{\textrm{erg}}(W) $. \end{proof}
\begin{theorem}[Coding theorem: weak converse]\label{weak-converse}
Let $W:A^{\zz}\times \mathcal{B}^{\zz}\to\cc$ be a stationary IMC channel. Then for each code $\mathcal{C}_{n}=(u_i,b_i)_{i=1}^{M_n}$ with $M_n\ge 2^{n(C_{\textrm{Holevo}}(W)+\eps)}$ we have 
\[\bar{e}(\mathcal{C}_n)\ge 1-\frac{C_{\textrm{Holevo}}(W)+\frac{1}{n} }{C_{\textrm{Holevo}}(W)+\eps},\]
where $\bar{e}(\mathcal{C}_n) $ denotes the average error probability of the code $\mathcal{C}_n$.
\end{theorem}
\begin{proof} This is an easy consequence of Fano inequality and Holevo bound and the proof is identical to that in the memoryless case. For completeness we provide the full argument; We may suppose that $\sum_{i=1}^{M_n}b_i=\idn$ since otherwise we can add $b_{M_n +1}:=\idn-\sum_{i=1}^{M_n}b_i$ to $b_i$ without affecting code performance. We define a stochastic matrix by setting
\[ K(j|i):=\textrm{tr}(D_{u_i}b_j)\quad (i=1,\ldots M_n, j=1,\ldots , M_n).\]
Consider the probability distribution $p$ on $A^n$ which assigns probability $\frac{1}{M_n}$ to each of the code words $u_i$. Then by Holevo bound \cite{holevo-1} we have
\[ \chi(p,W^n)\ge I(p,K),\]
where $I(p,K)$ denotes the mutual information computed with respect to given input and channel data $p$ and $K$. Combining this inequality with definition of the Holevo capacity (\ref{holevo-cap-def}) and with Fano inequality (see e.g. \cite{wolfowitz}) we obtain
\begin{eqnarray*}
 C_{\textrm{Holevo}}(W)&=&\sup_{l\in\nn}\frac{1}{l}C_{l}(W)\quad(\textrm{by eq. }(\ref{holevo-cap-def}))\\
&\ge & \frac{1}{n}C_n(W)\ge\frac{1}{n} I(p,K)\\
&\ge & \frac{1}{n}(  \log M_n -1- \bar{e}(\mathcal{C}_n)\log (M_n-1))\\
&\ge& (1-\bar{e}(\mathcal{C}_n))\frac{1}{n}\log (2^{n(C_{\textrm{Holevo}}(W)+\eps)})-\frac{1}{n}.
\end{eqnarray*}
This yields
\[\bar{e}(\mathcal{C}_n)\ge 1-\frac{C_{\textrm{Holevo}}(W)+\frac{1}{n} }{C_{\textrm{Holevo}}(W)+\eps},\]
as desired.\end{proof}
\section{Extension to DIMA Channels}\label{coding-dima}
The main ingredient in our proof of the direct part of coding theorem for IMC channels, Theorem \ref{coding-th-direct}, was Lemma \ref{prob-bounds} on probability bounds. There we heavily used our assumption that the chanel was IMC. If, instead of IMC condition, we merely assume that the channel is stationary ergodic we obtain, by an inspection of the proof of Lemma \ref{prob-bounds}, corresponding probability bounds for the induced channel
\begin{equation}\label{ind-channel}
W'^{n}(x^n,b):=\frac{\psi_{p,W}(|x^n\rangle\langle x^n|\otimes b)}{p^n (x^n)}\quad (b\in\mathcal{B}^{n}),
\end{equation}
which is defined for all $x^n\in A^{n}$ with $p^n (x^n)\neq 0$ for a fixed stationary ergodic $p\in \mathcal{P}(A^{\zz},\Sigma_c)$. Indeed, we may w.l.o.g. restrict ourselves to those $x^n \in A^{n}$ with $p^n (x^n)=p([x^n])\neq 0$, where $[x^n]$ denotes the cylinder set generated by $x^n$. Then observe that
\begin{equation}\label{ind-channel-integral}
W'^{n}(x^n,b)=\frac{1}{p^{n}(x^n)}\int_{[x^n]}W(x,b)p(dx) \quad (b\in \mathcal{B}^{n} ),
\end{equation}
by our definition of the joint state, eq. (\ref{joint-state}).
Now, it suffices to replace $W^n$ by $W'^n$ in the proof of Lemma \ref{prob-bounds} to obtain the desired extension of that result. Note that the code construction in Theorem \ref{coding-th-direct} depends only on Lemma \ref{prob-bounds}, thus we obtain for all $\delta>0$, $\lambda\in (0,1)$ and all sufficiently large $n \in \nn$  codes $\mathcal{C}_{n}=(u_i,b_i)_{i=1}^{M_n}$ with
\begin{enumerate}
\item $\max_{i\in \{1,\ldots , M_n \}}(1-W'^n(u_i,b_i))\le \lambda$
\item $M_n\ge 2^{n(C_{\textrm{erg}}(W)-\delta)}$,
\end{enumerate}
where $C_{\textrm{erg}}(W) $ is now defined as
\begin{equation}\label{erg-cap-dima}
C_{\textrm{erg}}(W):=\sup_{p \textrm{ stationary ergodic}} i(p,W),
\end{equation}
with
\begin{equation}\label{mut-inf-dima}
 i(p,W):=s(\psi_p) + s(\psi_{q(p)})-s(\psi_{p,W}). 
\end{equation}
Since $W'^n$ appears in the first item above, the code we have obtained is not a code for the channel $W$ with prescribed error probability. If we assume in addition to stationarity and ergodicity that $W$ fulfills DIMA condition , eq. (\ref{dima-def}), the code above is easily converted into one with low error probability for $W$: By eq. (\ref{ind-channel-integral}) we have
\[1-\lambda\le W'^{n}(u_i,b_i)=\frac{1}{p^{n}(u_i)}\int_{[u_i]}W(x,b_i)p(dx),\]
for all $i\in \{1,\ldots ,M_n\}$ so that for each $i$ there must be at least one $x(i)\in A^{\zz}$ with
\[ W(x(i), b_i)\ge 1-\sqrt{\lambda}.\]
Employing the DIMA condition we find positive integers $m,a$ such that
\[|W(x,b_i)-W(x(i),b_i)|\le \sqrt{\lambda}, \]
for all $i\in \{1,\ldots ,M_n\}$ whenever $x_j=x(i)_j$ for $1-m\le j\le n+a+1$. Thus setting $u'_i=x(i)_{1-m}^{n+a+1}$ and $b'_i:=b_i \in \mathcal{B}^{n}\subset \mathcal{B}^{[1-m,n+a-1]}$ we obtain the desired sequence of codes for the channel $W$ after shifting the sequences $u'_i$ as well as the decoding operators $b'_i$ $m$ places to the right. Thus we have proved
\begin{theorem}[DIMA Coding theorem: direct part]\label{dima-direct-part}
Let $W: A^{\zz}\times \mathcal{B}^{\zz}\to\cc$ be a stationary ergodic DIMA cq-channel. Then $C_{\textrm{erg}}(W) $, defined in eq. (\ref{erg-cap-dima}), is an achievable rate, and thus $C(W)\ge C_{\textrm{erg}}(W) $. 
\end{theorem}
In the proof of the converse part we shall need the periodic product information capacity which is defined by
\begin{equation}\label{per-prod-cap}
  C_{pp}(W):=\sup_{p \textrm{ periodic product}}i(p,W),
\end{equation}
$i(p,W)$ being given by eq. (\ref{holevo-cap-def}).
\begin{theorem}[DIMA coding theorem: weak converse]\label{weak-converse-dima}
Let $W: A^{\zz}\times \mathcal{B}^{\zz}\to\cc$ be a stationary ergodic DIMA cq-channel. Then $C_{pp}(W)=C_{\textrm{erg}}(W)$ holds. Moreover, for each $\eps>0$ and any code $\mathcal{C}_n =(u_i,b_i)_{i=1}^{M_n}$ with $M_n\ge 2^{n(C_{pp}(W)+\eps )}$ we have
\[  \bar{e}(\mathcal{C}_n) \ge 1-\frac{C_{pp}(W)+\frac{1}{n}}{C_{pp}(W)+\eps }   . \]
\end{theorem}
\begin{proof} We divide the proof in two parts. In the first part we infer from Holevo bound and Fano's inequality that $C_{pp}(W)\ge C(W)$ holds for each stationary cq-channel $W$. For DIMA channels we then know that $C_{pp}(W)\ge C(W)\ge C_{\textrm{erg}}(W)$ from the direct part of the coding theorem, Theorem \ref{dima-direct-part}. Finally we show that $C_{pp}(W)=C_{\textrm{erg}}(W) $ which is a consequence of the affinity of the v. Neumann entropy rate on periodic states and the fact that shifting a periodic state one site to the left/right does not change its entropy rate.\\
Let $\mathcal{C}_n =(u_i, b_i)_{i=1}^{M_n}$ be a code for $W$ with $M_{n}\ge 2^{n(C_{pp}(W)+\eps)}$. Let $p_n$ denote the distribution on $A^n$ which assigns probability $\frac{1}{M_n}$ to each of the code words $u_i$, $i=1,\ldots ,M_n$ and consider the product probability measure $p:=\cdots p_n\otimes p_n\otimes p_n\cdots$ with period $n$ on $(A^{\zz},\Sigma_c)$. Then we have $h(p)=\frac{1}{n}\log M_n$. Moreover, we need the family of induced channels $\{W'^{l}\}_{l\in\nn}$ from eq. (\ref{ind-channel}), or equivalently eq. (\ref{ind-channel-integral}), defined with respect to the periodic product measure $p$. If we denote by $D_{\psi_{p,W}}^l$ the density operator of the state $\psi_{p,W}^l$, then we have
\[D_{\psi_{p,W}}^l={\sum_{x^l\in A^l}}'p^l (x^l)|x^l\rangle\langle x^l|\otimes D_{x^l}, \]
where $ D_{x^l}$ is the density operator of $W'^{l}(x^l,\cdot )$ and ${\sum}'$ indicates that the summation is performed only over those $x^l$ with $p^{l}(x^l)>0$.
Note that then for each $l\in \nn$
\begin{equation}\label{holevo-dima-1}
  \chi(p^l, W'^{l})=S(\psi_{p}^{l})+S(\psi_{q(p)}^{l})-S(\psi_{p,W}^{l})
\end{equation}
holds, and thus we have
\begin{equation}\label{holevo-dima-12}
  i(p,W)=\lim_{l\to\infty}\frac{1}{l}  \chi(p^l, W'^{l}),
\end{equation}
by eq. (\ref{mut-inf-dima}). If we introduce the average error probability $\bar{e}(\mathcal{C}_{n},W')$ with respect to the induced channel $W'$, i.e.
\begin{equation}\label{average-error-w'}
  \bar{e}(\mathcal{C}_{n},W'):=\frac{1}{M_n}\sum_{i=1}^{M_n}(1-W'^{n}(u_i,b_i)),
\end{equation}
 then we can argue in a similar fashion as in the proof of Theorem \ref{weak-converse} and obtain
\begin{equation}\label{holevo-dima-2}
 \chi(p_n,W'^{n})\ge (1-\bar{e}(\mathcal{C}_n,W'))\log (2^{n(C_{pp}(W)+\eps)})-1.
\end{equation}
For each $l\in\nn$ we write $l=kn+r$, $0\le r<n$, and using superadditivity of $\chi$ with respect to product states together with the $n-$periodicity of the channel and resulting states, which follows from the subadditivity of the conditional v. Neumann entropy (see Theorem 11.16 in \cite{nielsen-chuang}), we arrive at
\begin{equation}\label{holevo-dima-3}
 \chi(p^{l},W'^{l})\ge k \chi(p_n,W'^{n})+ \chi(p_n^{[kn+1,kn+r]}, W'^{[kn+1,kn+r]}). 
\end{equation}
Combining eq. (\ref{holevo-dima-3}), eq. (\ref{holevo-dima-2}) and dividing by $l$ we obtain
\begin{eqnarray}
  \frac{1}{l} \chi(p^{l},W'^{l})&\ge & \frac{k}{l}\chi(p_n,W'^{n})+o(1)\nonumber\\
&\ge & \frac{k}{l}((1-\bar{e}(\mathcal{C}_n,W'))\log (2^{n(C_{pp}(W)+\eps)})-1 )\nonumber\\
& & +o(1)\nonumber\\
&=&  \frac{k}{l}((1-\bar{e}(\mathcal{C}_n,W'))n(C_{pp}(W)+\eps)-1 )\nonumber \\
& & + o(1).\nonumber
\end{eqnarray}
Taking the limit $l\to\infty$, taking into account eq. (\ref{holevo-dima-1}), eq. (\ref{holevo-dima-12}), definition of $C_{pp}(W)$, eq. (\ref{per-prod-cap}), leads to
\begin{eqnarray}\label{holevo-dima-4}
  C_{pp}(W)&\ge & i(p,W)\nonumber\\
&\ge & (1-\bar{e}(\mathcal{C}_n,W'))(C_{pp}(W)+\eps)-\frac{1}{n},
\end{eqnarray}
or, equivalently
\begin{equation}\label{holevo-dima-5}
 \bar{e}(\mathcal{C}_n,W')\ge 1-\frac{C_{pp}(W)+\frac{1}{n}}{C_{pp}(W)+\eps }.  
\end{equation}
From definition of average error probability, eq. (\ref{average-error}), we infer that $\bar{e}(\mathcal{C}_n)\ge \bar{e}(\mathcal{C}_n,W') $ and thus inequality (\ref{holevo-dima-5}) yields
\begin{equation}\label{holevo-dima-6}
  \bar{e}(\mathcal{C}_n)\ge \bar{e}(\mathcal{C}_n,W')\ge 1-\frac{C_{pp}(W)+\frac{1}{n}}{C_{pp}(W)+\eps }.
\end{equation}
This shows that for each stationary cq-channel $W$ the inequality
\begin{equation}\label{holevo-dima-7}
  C_{pp}(W)\ge C(W)
\end{equation}
holds. If $W$ is in addition DIMA, then the direct part, Theorem \ref{dima-direct-part}, yields with eq. (\ref{holevo-dima-7})
\[ C_{pp}(W)\ge C(W)\ge C_{\textrm{erg}}(W).\]
Thus we have to prove the converse inequality
\begin{equation}\label{holevo-dima-8}
  C_{pp}(W)\le C_{\textrm{erg}}(W).
\end{equation}
For any $\delta>0$ there is a product probability measure $p$ on $(A^{\zz},\Sigma_{c})$ with period $t\in\nn$ with
\begin{equation}\label{holevo-dima-9}
  C_{pp}(W)-\delta\le i(p,W).
\end{equation}
Then the probability measure given by
\[p':=\frac{1}{t}\sum_{i=0}^{t-1}p\circ T_{in}^{-i}\]
is stationary ergodic. Note that the joint input-output state and output state depend affinely on the input probability measure. Using the defining formula (\ref{joint-state-2}) for joint input-output state and resulting output state together with change of variable formula one immediately sees that
\[ \psi_{p\circ T_{in}^{-i},W}(f\otimes b)=\psi_{p,W}(T_{in}^{i}f\otimes T_{out}^{i}b),\]
and
\[\psi_{q(p\circ T_{in}^{-i})}(b)=\psi_{q(p)}(T_{out}^{i}b)\]
hold. Arguing as in the proof of Theorem 3.1.3 in \cite{bksss} shows that
\[s(\psi_{p\circ T_{in}^{-i},W} )=s(\psi_{p,W}),\quad s(\psi_{q(p\circ T_{in}^{-i})} )=s(\psi_{q(p)}),\] 
and
\[s(p\circ T_{in}^{-i})=s(p).\]
Using this and the affinity of the v. Neumann entropy rate on periodic states, we obtain
\[ i(p',W)=\frac{1}{t}\sum_{i=0}^{t-1}i(p\circ T_{in}^{-i},W)=i(p,W).\]
This inequality together with (\ref{holevo-dima-9}) shows that inequality (\ref{holevo-dima-8}) is valid since $\delta>0$ was arbitrary. \end{proof}
\section{Classical Capacity of Output Weakly Mixing Quantum Channels}\label{coding-class-quant}
In this section we show that the results obtained so far imply immediately capacity results for the transmission of classical information through output weakly mixing quantum channels. The extension to ergodic quantum channels is postponed to the future work in order to keep the size of this paper reasonable.\\
We consider two finite-dimensional Hilbert spaces $\hr_1$, $\hr_2$ and corresponding algebras of linear operators $\mathcal{A}:=\mathcal{L}(\hr_1)$, $\mathcal{B}:=\mathcal{L}(\hr_2)$ together with quasi-local $C^{\ast}-$algebras $\mathcal{A}^{\zz}$ and $\mathcal{B}^{\zz}$. A quantum channel is described by a linear, completely positive, unital map $K:\mathcal{B}^{\zz}\to \mathcal{A}^{\zz}$. $K$ is called $(T_{in},T_{out})-$stationary if $K\circ T_{out}=T_{in}\circ K$ holds for the shifts $T_{out}$ resp. $T_{in}$ on $\mathcal{B}^{\zz}$ resp. $\mathcal{A}^{\zz}$. The quantum channel $K$ is $(T_{in},T_{out})-$\emph{ergodic} if it is extreme point in the convex set of $(T_{in},T_{out})-$stationary quantum channels. \\
A convenient sufficient condition for ergodicity of a stationary channel $K:\mathcal{B}^{\zz}\to \mathcal{A}^{\zz}$ , as in the classical theory (cf. \cite{gray}), is that it is output weakly mixing: A quantum channel $K$ is said to be \emph{output weakly mixing} if for any state $\varphi \in \mathcal{S}(\mathcal{A}^{\zz})$ and all $b_1,b_2\in\mathcal{B}^{\zz}$
\begin{equation}\label{output-weakly-mixing-def}
  \lim_{n\to\infty}\frac{1}{n}\sum_{i=0}^{n-1}|\varphi (K(b_1 T_{out}^{i}b_2))-\varphi (K(b_1)K (T_{out}^{i}b_2))|=0
\end{equation}
holds. Obviously, the condition in eq. (\ref{output-weakly-mixing-def}) need only to be checked on elements in $\mathcal{B}^{loc}$ (see section \ref{quasi-local-app}). The proof that each output weakly mixing channel is ergodic will be given in a forthcoming paper. We will merely show here how such a channel induces an ergodic cq-channel.
\begin{remark}\label{rel-to-class-out-mixung}
It is readily seen that the condition (\ref{output-weakly-mixing-def}) for cq-channels is equivalent to
\[ \lim_{n\to\infty}\frac{1}{n}\sum_{i=0}^{n-1}|W(x,b_1 T_{out}^{i}b_2 )-W(x,b_1)W(x,T_{out}^{i}b_2)|=0\]
for all $x\in A^{\zz}$, which reflects the classical condition of a channel of being output weakly mixing as given in \cite{gray}.
\end{remark}
The continuity notions of section \ref{continuity} are easily extended to the present setting. E.g. channel $K:\mathcal{B}^{\zz}\to\mathcal{A}^{\zz}$ is said to have \emph{decaying input memory and anticipation} (DIMA) if for all $n\in \zz$, $k\in \nn$ 
\[\lim_{m,a\to\infty}\sup_{\varphi,\varphi'\in \mathcal{S}(\mathcal{A}^{\zz}):\varphi=_{m,a,n,k}\varphi'}d_{n,k}(\varphi\circ K,\varphi'\circ K)=0, \]
where
\[d_{n,k}(\varphi\circ K,\varphi'\circ K):=\sup_{b\in\mathcal{B}^{[n,n+k]}:0\le b\le \idn}|\varphi\circ K(b)-\varphi'\circ K(b)|, \]
and $\varphi=_{m,a,n,k}\varphi'$ means that the states $\varphi$ and $\varphi'$ on $\mathcal{A}^{\zz}$ have equal restrictions to $\mathcal{A}^{[n-m,n+k+a]}$.\\
A quantum channel $K:\mathcal{B}^{\zz}\to\mathcal{A}^{\zz}$ is called \emph{input memoryless and causal} (IMC) if $K(\mathcal{B}^{[n,n+k]})\subset \mathcal{A}^{[n,n+k]}$ for all integers $n$ and $k$ with $k\ge 0$.\\
We denote, as before, $\mathcal{A}^{[1,n]}$ by $\mathcal{A}^n$ with a similar abbreviation for $\mathcal{B}^{[1,n]}$.
A \emph{code of length $M_n$ for transmission of classical information} via quantum channel $K:\mathcal{B}^{\zz}\to\mathcal{A}^{\zz}$ is a family $\mathcal{C}_n:=(\varphi_i,b_i)_{i=1}^{M_n}$ consisting of states $\varphi_i\in\mathcal{S}(\mathcal{A}^n)$, $i=1,\ldots , M_n$, and decoding operators $0\le b_i\in \mathcal{B}^{n}$ with $\sum_{i=1}^{M_n}b_i\le \idn$. \\
The \emph{error probability} of a code $\mathcal{C}_n$ is given by
\begin{equation}
 e(\mathcal{C}_n):=\max_{i\in \{1,\ldots,M_n\}}\sup_{\bar{\varphi}_i\in\mathcal{S}(\mathcal{A}^{\zz}):\bar{\varphi}_i^{n}=\varphi_i}(1-\bar{\varphi}_i(K(b_i))), 
\end{equation}
where $\bar{\varphi}_i^{n} $ denotes the restriction of $\bar{\varphi}_i$ to $\mathcal{A}^n$. The \emph{capacity} of the channel $K$ is then defined in the usual way.\\
We consider a stationary output weakly mixing IMC quantum channel $K:\mathcal{B}^{\zz}\to\mathcal{A}^{\zz}$. Our goal is to show that the classical capacity of this channel is given by
\[ C_{\textrm{Holevo}}(K):=\lim_{n\to\infty}\frac{1}{n}C_n(K),\]
where
\[ C_n(K):=\sup_{\{p_i,\varphi_i\}_{i=1}^{m}}\left(S\left(\sum_{i=1}^{m}p_i\varphi_i\circ K \right)-\sum_{i=1}^{m}p_iS(\varphi_i\circ K)\right)\]
and the least upper bound is taken over all ensembles on $\mathcal{A}^{n}$, i.e. $p_i\ge 0$ with $\sum_{i=1}^m p_i=1$ and each $\varphi_i$ is a state on $\mathcal{A}^{n}$. For any $\eta >0$ there is a positive integer $n$ and an ensemble $\{p_i,\varphi_i\}_{i=1}^m $such that 
\begin{equation}\label{last-1}
S\left(\sum_{i=1}^{m}p_i\varphi_i\circ K \right)-\sum_{i=1}^{m}p_iS(\varphi_i\circ K)\ge C_{\textrm{Holevo}}(K)-\eta  
\end{equation}
holds. Let $A:=\{1,2,\ldots ,m\}$ and for each $x\in A^{\zz}$ consider the states $\varphi_x:=\ldots \otimes \varphi_{x_{-1}}\otimes \varphi_{x_{0}}\otimes \varphi_{x_{1}}\otimes \ldots$ on $\mathcal{A}^{\zz}$. Moreover we consider the stationary product distribution $p$ built up from the probability vector $(p_i)_{i=1}^{m}$. Let $W:A^{\zz}\times \mathcal{B}^{\zz}\to\cc$ be the cq-channel given by
\[ W(x,b):=\varphi_{x}(K(b)).\]
It is readily seen using our assumption that $K$ is stationary and output weakly mixing that the cq-channel $W$ is stationary, IMC and output weakly mixing. A calculation similar to the proof of Lemma 9.4.3 in \cite{gray} shows that $W$ is ergodic. Thus all results from section \ref{coding-imc} apply and show that by Theorem \ref{coding-th-direct} there is a sequence of codes $\mathcal{C}_n$ of lengths $M_n$ for the cq-channel $W$ with
\[ \liminf_{n\to\infty }\frac{1}{n}\log M_n\ge C_{\textrm{Holevo}}(W),\]
where $C_{\textrm{Holevo}}(W)$ is defined in eq. (\ref{holevo-cap-def}). It is obvious that the codes $\mathcal{C}_n$ for $W$ generate codes $\mathcal{C'}_n$ for $K$ with the same lengths and error probabilities. Note that
\[ \chi (p^1,W^1)=S\left(\sum_{i=1}^{m}p_i\varphi_i\circ K \right)-\sum_{i=1}^{m}p_iS(\varphi_i\circ K),\]
And thus from (\ref{holevo-cap-def}) and (\ref{last-1}) we can infer that
\[\liminf_{n\to\infty }\frac{1}{n}\log M_n\ge C_{\textrm{Holevo}}(K)-\eta  \]
holds. This shows that all rates  below $C_{\textrm{Holevo}}(K) $ are achievable. The weak converse is shown in the same vein as in the memoryless case, see the proof of Theorem \ref{weak-converse}.
\section*{Acknowledgment}
IB wishes to thank Rainer Siegmund-Schultze for introducing him into the theory of classical communication channels and several discussions on various aspects of this theory over the last years. \\
Support and funding from the Deutsche Forschungsgemeinschaft (DFG) via project Bj 57/1-1 ``Entropie und Kodierung gro\ss er Quanten-Informationssysteme'' is gratefully acknowledged.
\section{Appendix}\label{appendix}
\subsection{$C^{\ast}-$Algebras, States and $\zz$-Actions}\label{c-ast-app}
$C^{\ast}-$ algebras are axiomatic generalizations of well known objects, such as the continuous functions over compact spaces or bounded operators over finite- or infinite-dimensional Hilbert spaces, which, additionally to their algebraic structure given by possibility to add and multiply elements, have an adjoint operation and norm defined on it. Excellent introduction to basic concepts and methods relevant for applications of $C^{\ast}-$algebras is given in \cite{kadison-1}.\\
Let $\mathcal{A}$ be a linear space over the field $\cc$ which is additionally endowed with a  distributive and associative product. An adjoint operation $^{\ast}:\mathcal{A}\to\mathcal{A}$  is an anti-linear map (i.e. $(\lambda a+\mu a')^{\ast}=\bar{\lambda}a^{\ast}+\bar{\mu}a'^{\ast}$ ) with $a^{\ast \ast}=a$ and $(aa')^{\ast}=a'^{\ast}a^{\ast}$ for $a,a'\in\mathcal{A},\lambda,\mu\in\cc$. An algebra $\mathcal{A}$ over $\cc$ equipped with a adjoint operation is called $^{\ast}-$\emph{algebra}.\\
A $^{\ast}-$algebra $\mathcal{A}$ is a $C^{\ast}-$\emph{algebra} if there is a norm $||\cdot ||:\mathcal{A}\to[0,\infty)$ such that
\begin{enumerate}
\item $\mathcal{A}$ is complete with respect to $||\cdot ||$.
\item $||aa'||\le ||a||\cdot ||a'||$ for all $a,a'\in\mathcal{A}$.
\item $||a^{\ast}a||=||a||^{2}$ for all $a\in\mathcal{A}$.
\end{enumerate}
Standard examples of $C^{\ast}-$algebras are:
\begin{enumerate}
\item The set $C(X)$ of continuous $\cc -$valued functions on a compact Hausdorff space equipped with the sup-norm $||\cdot||_{\infty}$. The adjoint operation is given by complex-conjugation of functions.
\item The set $\mathcal{B}(\hr)$ of bounded operators acting on a Hilbert space with the operator norm. $^{\ast}$ is then the usual adjoint operation.
\item Quasi-local algebras in the next section \ref{quasi-local-app}.
\end{enumerate}
A $C^{\ast}-$ algebra $\mathcal{A}$ is called \emph{unital} if there is an element $\idn\in \mathcal{A}$, called the identity, with $\idn a=a$ for all $a\in \mathcal{A}$. In this paper we will be concerned only with unital algebras.\\
A state on a unital $C^{\ast}-$ algebra $\mathcal{A}$ is a $\cc -$linear functional $\psi:\mathcal{A}\to \cc$ with 
\begin{enumerate}
\item $\psi (a)\ge 0$ for all $a\in \mathcal{A}$ with $a\ge 0$. Here $a\ge 0$ means that there is $b\in\mathcal{A}$ with $a=b^{\ast}b$.
\item $\psi (\idn)=1$. 
\end{enumerate}
For a compact Hausdorff set $X$, states on $(C(X), ||\cdot ||_{\infty})$ can be uniquely associated to probability measures on $(X,\Sigma_{Borel})$ via Riesz-Markov representation theorem (see \cite{rudin} Theorem 2.14).\\
If we consider a finite-dimensional Hilbert space $\hr$ and $\mathcal{A}=\mathcal{B}(\hr)$ then using the Hilbert-Schmidt inner product $\langle a,b\rangle =\textrm{tr}(a^{\ast}b)$, $a,b\in\mathcal{A}$, and Riesz representation theorem from elementary linear algebra it is easily seen that each state $\psi$ on $\mathcal{A}$ can be represented by a unique density operator $D\in \mathcal{A}$ (i.e. $D=D^{\ast}$, $D\ge 0$ and $\textrm{tr}(D)=1$), i.e
\[\psi (a)=\textrm{tr}(Da)\quad \forall a\in\mathcal{A}.\]
A $^{\ast}-$\emph{automorphism} of a $C^{\ast}-$algebra is a one-to-one, onto linear map $T:\mathcal{A}\to\mathcal{A}$ with
$T(a^{\ast})=(T(a))^{\ast}$ and $T(aa')=T(a)T(a')$ for all $a,a' \in \mathcal{A}$. Any $^{\ast}$-automorphism induces a $\zz -$\emph{action} on $\mathcal{A}$, i.e. a family $(T_{z})_{z\in\zz}$ of $^{\ast}-$automorphisms with $T_{z_1+z_2}=T_{z_1}\circ T_{z_2}$ and $T_{0}=\textrm{id}_{\mathcal{A}}$. This family is given by $T_{z}:=T^{z}$, $z\in\zz$. Conversely, each $\zz -$action is given in this way, simply set $T:=T_{1}$.\\
A state $\psi$ on $\mathcal{A}$ is $\zz -$\emph{invariant}, or, equivalently, $T-$\emph{invariant}, if $\psi\circ T=\psi$ holds. It is obvious that $\psi\circ T_z=\psi$ for all $z\in \zz$. The set of $\zz -$invariant states is convex. An $\zz -$invariant state $\psi$ is called \emph{ergodic} if $\psi$ is an extreme point in the convex set of $\zz -$invariant states. Somewhat more concrete example of a $\zz -$action and a necessary and sufficient criterion for the ergodicity, that parallels the classical setting, can be found in the next section \ref{quasi-local-app}.
\subsection{Quasi-Local Algebras and Ergodic States}\label{quasi-local-app}
Quasi-local algebras are used to describe interacting systems of infinitely many spins over a lattice $\zz^{d}$ in quantum statistical mechanics. We will consider only the case $d=1$, but constructions generalize immediately to arbitrary dimension. A readable introduction to quasi-local algebras and ergodic states on such algebras is given in \cite{simon}.\\
Let us consider a finite-dimensional $C^{\ast}-$algebra $\mathcal{A}$. For definiteness, let us consider the case that either $\mathcal{A}=\mathcal{L}(\hr)$ is the $C^{\ast}-$algebra of linear operators acting on some finite-dimensional Hilbert space $\hr$, or $\mathcal{A}=\mathfrak{F}(A)$, where $\mathfrak{F}(A)$ denotes the set of all $\cc -$valued functions on a finite set $A$. These two examples will suffice for our concerns.
Suppose that to each $n\in\zz$ we attach a copy $\mathcal{A}_n$ of $\mathcal{A}$. Let $\Lambda\subset \zz$ be a finite set and let us define $\mathcal{A}^{\Lambda}:=\bigotimes_{n\in \Lambda}\mathcal{A}_n$. $\mathcal{A}^{\Lambda} $ is called the algebra of observables belonging to sites in $\Lambda$. For $\Lambda\subset\Lambda'\subset\zz$, both finite, there is a natural embedding of $\mathcal{A}^{\Lambda}$ into $\mathcal{A}^{\Lambda'}$ given by 
\[\mathcal{A}^{\Lambda}\ni a\mapsto a\otimes \idn_{\Lambda'\setminus\Lambda}\in \mathcal{A}^{\Lambda'}, \]
where $\idn_{\Lambda'\setminus\Lambda} $ denotes the identity in $\mathcal{A}^{\Lambda'\setminus\Lambda }$. Note also that 
\[ ||a\otimes \idn_{\Lambda'\setminus\Lambda}||_{\mathcal{A}^{\Lambda'}}=||a||_{\mathcal{A}^{\Lambda}}\]
holds for all $a\in \mathcal{A}^{\Lambda}$.
Moreover, if for two finite subsets $\Lambda_1$, $\Lambda_2$ of $\zz$ we have $a\in\mathcal{A}^{\Lambda_1}$ and $a'\in\mathcal{A}^{\Lambda_2}$ then the product, adjoint operation, and other algebraic constructions can be naturally preformed in the larger algebra $\mathcal{A}^{\Lambda_1\cup \Lambda_2}$. Thus, if we set
\[ \mathcal{A}^{loc}:=\bigcup_{\Lambda\subset \zz: |\Lambda|<\infty}\mathcal{A}^{\Lambda}\]
we obtain the normed $^{\ast}-$algebra of \emph{local observables}. Its norm-completion
\[ \mathcal{A}^{\zz}:=\overline{\mathcal{A}^{loc} }\]
is then called the \emph{quasi-local} $C^{\ast}-$\emph{algebra} built up from $\mathcal{A}$.\\
For example, if $\mathcal{A}=\mathfrak{F}(A) $ then it is an immediate consequence of Stone-Weierstrass theorem (cf. \cite{kadison-1} theorem 3.4.14) that $\mathcal{A}^{\zz}=C(A^{\zz})$, the set of continuous functions on $A^{\zz}$ with $||\cdot ||_{\infty}-$norm, where $A^{\zz}$ is equipped with the product topology.
\begin{remark}\label{app-remark-1}
Note the similarity of the construction of quasi-local $C^{\ast}$-algebra to the construction of $\sigma$-algebra $\Sigma_c$ on space of doubly-infinite sequences drawn from some finite alphabet $A$ (cf. \cite{shields}): For the latter purpose one starts for each $n\in \nn$ with algebra $\Sigma_n$ of sets which is generated by the cylinder sets with the base in $A^{[-n,n]}$ and observes that $\Sigma_n\subset \Sigma_{n+1}$. Then it is clear that $\Sigma_{loc}:=\bigcup_{n\in\nn}\Sigma_n$ is an algebra of sets. Then $\Sigma_c$ is simply defined by $\Sigma_c:=\sigma(\Sigma_{loc})$, i.e. as the $\sigma$-algebra generated by $\Sigma_{loc}$. The main difference to the quasi-local setting lies in the kind of approximation. In the quasi-local algebra each observable can be approximated uniformly by local ones, whereas the approximation in $\Sigma_c$ means that for each probability measure $p$ on $(A^{\zz},\Sigma_c)$ and each $A\in \Sigma_c$ there is a sequence $(A_n)_{n\in\nn}$ in $\Sigma_{loc}$ with $\lim_{n\to\infty}p(A\triangle A_n )=0$. Here $\triangle$ denotes the symmetric difference of the sets.
\end{remark}
Any state $\psi$ on $\mathcal{A}^{\zz}$ induces a family of states $(\psi^{\Lambda})_{\Lambda\subset\zz:|\Lambda|<\infty}$ on $\mathcal{A}^{\Lambda}$. $\Lambda\subset \Lambda'$ implies $\psi^{\Lambda'}\upharpoonright \mathcal{A}^{\Lambda}=\psi^{\Lambda}$, i.e. the family $(\psi^{\Lambda})_{\Lambda\subset\zz:|\Lambda|<\infty} $ is \emph{consistent}. Conversely, any consistent family of states $(\psi^{\Lambda})_{\Lambda\subset\zz:|\Lambda|<\infty} $ defines a state $\psi$ on $\mathcal{A}^{\zz}$.
\begin{remark}
At this point we have again a nice analogy to the classical case. According to Kolmogorov's consistency theorem (cf. \cite{shields} for a concise discussion) any probability measure $p$ on $(A^{\zz},\Sigma_c)$ is uniquely determined by (or can be constructed from) the set of its finite-dimensional marginal distributions on $A^{[n,m]}$, $n\le m, n,m\in \zz$.
\end{remark}
A shift $T:\mathcal{A}^{\zz}\to\mathcal{A}^{\zz}$ on $\mathcal{A}^{\zz}$ is induced by 
\[ \mathcal{A}^{\Lambda}\ni a\simeq a\otimes\idn_{\mathcal{A}}\mapsto T(a):=\idn_{\mathcal{A}}\otimes a\simeq a\in \mathcal{A}^{\Lambda+1}. \]
Note that the shift $T$ is a $^{\ast}-$isomorphism, i.e. it is linear, fulfills $T(a^{\ast})=(T(a))^{\ast}$ and $T(ab)=T(a)T(b)$ for all $a,b\in\mathcal{A}^{\zz}$.\\
A state $\psi$ on $\mathcal{A}^{\zz}$ is called \emph{stationary} if $\psi\circ T=\psi$ holds. The set of stationary states on $\mathcal{A}^{\zz}$ is convex. A state $\psi$ on $\mathcal{A}^{\zz}$ is called \emph{ergodic} if it is an extreme point in the set of stationary states. It can be shown (cf. \cite{simon} theorem 1.7.10) that, for quasi-local algebras and shifts, the statement that $\psi$ is ergodic is equivalent to
\begin{equation}\label{ergodic-characterization}
\lim_{n\to\infty}\frac{1}{2n+1}\sum_{i=-n}^{n}\psi(aT^{i}(b))=\psi(a)\psi(b) 
\end{equation}
for all $a,b\in\mathcal{A}^{\zz}$.\\
The v. Neumann entropy rate of a stationary state $\psi$ on $\mathcal{A}^{\zz}$ is given by
\begin{equation}\label{vn-entropy-def}
s(\psi):=\lim_{n\to\infty}\frac{1}{n}S(\psi^n), 
\end{equation}
where $\psi^n:=\psi\upharpoonright \mathcal{A}^{[1,n]}$, $[1,n]:=\{1,2,\ldots ,n\}$, and
\[ S(\psi^n):= -\textrm{tr}(D_{\psi^n}\log D_{\psi^n})\]
is the v. Neumann entropy.
$D_{\psi^n}\in\mathcal{A}^{[1,n]}$ denotes the density operator of $\psi^n$. Note that the limit in eq. (\ref{vn-entropy-def}) exist and equals $\inf_{n\in\nn}\frac{1}{n}S(\psi^n)$ since v. Neumann entropy is subadditive and $\psi$ is assumed stationary;
\[ S(\psi^{n+m})\le S(\psi^n)+S(\psi^m). \]
For a proof of the last inequality see e.g. \cite{nielsen-chuang}.

\subsection{Completely Positive Maps and Quantum Channels}\label{comp-pos-app}
In this part of the appendix we provide the basic definition of complete positivity and make an attempt to explain how this fits to the notion of a channel from the classical information theory. The standard reference for completely positive maps is the monograph \cite{paulsen} by Paulsen.\\
Let $\mathcal{A},\mathcal{B}$ be $C^{\ast}-$algebras and consider a linear map $E:\mathcal{B}\to \mathcal{A}$. $E$ is called \emph{positive} if $E(b)\ge 0$ for all $b\in\mathcal{B}$ with $b\ge 0$. Suppose that $\mathcal{B},\mathcal{A}$ are unital, a linear map $E:\mathcal{B}\to \mathcal{A}$ is \emph{unital} if $E(\idn_{\mathcal{B}})=\idn_{\mathcal{A}}$. \\
Now, we consider additionally the $C^{\ast}-$algebra of $n$-by-$n$ complex matrices $\mathbb{M}(n,\cc)$ and $\mathcal{B}\otimes\mathbb{M}(n,\cc) $ respectively $\mathcal{A}\otimes\mathbb{M}(n,\cc) $, both of which can be endowed with a canonical structure of $C^{\ast}-$algebra (see \cite{paulsen} for details). Essentially, these $C^{\ast}-$structures are given by identifying the members of $\mathcal{B}\otimes\mathbb{M}(n,\cc) $ resp. $\mathcal{A}\otimes\mathbb{M}(n,\cc) $ with the $n$-by-$n$ matrices having entries from $\mathcal{B}$ resp. $\mathcal{A}$.\\
A linear map $E:\mathcal{B}\to \mathcal{A} $ is \emph{completely positive} if for each non-negative integer $n$ the map $E\otimes \textrm{id}_{\mathbb{M}(n,\cc) }:\mathcal{B}\otimes\mathbb{M}(n,\cc)\to \mathcal{A}\otimes\mathbb{M}(n,\cc)  $ is positive, where $\textrm{id}_{\mathbb{M}(n,\cc) } $ denotes the identity map on $\mathbb{M}(n,\cc) $.\\
A completely positive unital map $E:\mathcal{B}\to \mathcal{A} $ induces a map $E':\mathcal{S}(\mathcal{A})\to\mathcal{S}(\mathcal{B})$ between the sets of states via
\[E'(\psi):=\psi\circ E \quad (\psi \in\mathcal{S}(\mathcal{A}) ).\]
\emph{Quantum channels} are defined as completely positive, unital maps between $C^{\ast}$-algebras.
The connection to the classical channels is established via following result of Stinespring (see \cite{stinespring, paulsen}): If $\mathcal{B}$ or $\mathcal{A}$ is commutative (abelian) then each linear, positive map $E:\mathcal{B}\to \mathcal{A} $ is completely positive.\\
We will deal only with the simplest case in order to recover the classical channels. To this end let us consider two finite sets $Y$ and $X$ and a stochastic matrix $W:X\to Y$, i.e. $W(y|x)\ge 0$ and $\sum_{y\in Y}W(y|x)=1$ for all $x\in X$. Define $E:\mathfrak{F}(Y)\to \mathfrak{F}(X)$ by
\[ E(f)(x):=\sum_{y\in Y}f(y)W(y|x).\]
This map is obviously linear, positive and unital. And it is clear that each linear, positive and unital map is representable in this way. The induced map $E':\mathcal{P}(X)\to \mathcal{P}(Y)$ between the sets of probability distributions is then easily calculated:
\[ E'(p)(y)=\sum_{x\in X}p(x)W(y|x),\qquad (y\in Y)\]
which is exactly the output distribution of the channel for the stochastic input $p\in \mathcal{P}(X)$. This shows that classical channels fit nicely into the theory of completely positive maps.


\end{document}